\documentclass[12pt,preprint]{aastex}
\usepackage{emulateapj5}
\usepackage{apjfonts}
\usepackage{epsfig}
\usepackage{natbib}

\newcommand{\aox}{\ensuremath{\alpha_{\mathrm{ox}}}}

\newcommand{\chisq}{\ensuremath{\chi^2}}
\newcommand{\chisqr}{\ensuremath{\chi^2_r}}
\newcommand{\chandra}{\emph{Chandra}}

\newcommand{\etal}{et al.}

\newcommand{\feii}{\ion{Fe}{2}}

\def\gtrsim{\mathrel{\hbox{\rlap{\hbox{\lower4pt\hbox{$\sim$}}}\hbox{\raise2pt\hbox{$>$}}}}}
\newcommand{\fha}{\ensuremath{f_\mathrm{H{\alpha}}}}
\newcommand{\fwha}{\ensuremath{\mathrm{FWHM}_\mathrm{H{\alpha}}}}

\newcommand{\halpha}{H\ensuremath{\alpha}}

\newcommand{\hbeta}{H\ensuremath{\beta}}

\newcommand{\hn}{\halpha+[\ion{N}{2}]}
\newcommand{\hst}{\emph{HST}}

\newcommand{\kms}{km~s\ensuremath{^{-1}}}

\newcommand{\lf}{\ensuremath{L_{\rm{5100 \AA}}}}
\newcommand{\lum}{ergs s$^{-1}$}

\newcommand{\lledd}{\ensuremath{L_{\mathrm{bol}}/L{\mathrm{_{Edd}}}}}

\newcommand{\loiii}{\ensuremath{L_{\mathrm{[O {\tiny III}]}}}}
\newcommand{\lha}{\ensuremath{L_{\mathrm{H{\alpha}}}}}

\newcommand{\luv}{\ensuremath{L_{\rm{2500 \AA}}}}

\newcommand{\mbh}{\ensuremath{M_\mathrm{BH}}}

\newcommand{\mlb}{\ensuremath{M_{\mathrm{BH}}-L_{\mathrm{bulge}}}}
\newcommand{\msigma}{\ensuremath{M_{\mathrm{BH}}-\sigmastar}}
\newcommand{\msun}{\ensuremath{M_{\odot}}}

\newcommand{\nii}{[\ion{N}{2}]}

\newcommand{\oiii}{[\ion{O}{3}]}

\newcommand{\rosat}{\emph{ROSAT}}

\newcommand{\sii}{[\ion{S}{2}]}

\newcommand{\sigmastar}{\ensuremath{\sigma_{\ast}}}

\newcommand{\vc}{$v_\mathrm{c}$}
\newcommand{\vvmax}{$V/V_{\rm max}$}
\newcommand{\vmax}{$V_{\rm max}$}
\newcommand{\xmm}{{\it XMM-Newton}}

\def\lax{{$\mathrel{\hbox{\rlap{\hbox{\lower4pt\hbox{$\sim$}}}\hbox{$<$}}}$}}
\def\gax{{$\mathrel{\hbox{\rlap{\hbox{\lower4pt\hbox{$\sim$}}}\hbox{$>$}}}$}}

\slugcomment{To appear in {\it The Astrophysical Journal}.}
\shorttitle{BH Mass Function}
\shortauthors{GREENE \& HO}

\begin{document}

\title{The Mass Function of Active Black Holes in the Local Universe}

\author{Jenny E. Greene\altaffilmark{1}}
\affil{Department of Astrophysical Sciences, Princeton University, 
Princeton, NJ}
\altaffiltext{1}{Hubble and Princeton-Carnegie Fellow}

\author{Luis C. Ho}
\affil{The Observatories of the Carnegie Institution of Washington,
813 Santa Barbara St., Pasadena, CA 91101}

\begin{abstract}

We present the first measurement of the black hole (BH) mass function
for broad-line active galaxies in the local Universe.  Using the $\sim
9000$ broad-line active galaxies from the Fourth Data Release of the
Sloan Digital Sky Survey, we construct a broad-line luminosity
function that agrees very well with the local soft X-ray luminosity
function.  Using standard virial relations, we then convert observed
broad-line luminosities and widths into BH masses.  A mass function
constructed in this way has the unique capability to probe the mass
region $<10^6$~\msun, which, while insignificant in terms of total BH
mass density, nevertheless may place important constraints on the mass
distribution of seed BHs in the early Universe.  The characteristic
local active BH has a mass of $\sim 10^7$~\msun\ radiating at 10\% of 
the Eddington rate.  The active fraction
is a strong function of BH mass; at both higher and lower masses the
active mass function falls more steeply than one would infer from the
distribution of bulge luminosity.  The deficit of local massive
radiating BHs is a well-known phenomenon, 
while we present the first robust measurement of a decline in the space 
density of active BHs at low mass.
\end{abstract}


\section{The Local Black Hole Mass Function}

There is strong evolution in both the number density and typical
luminosity of active galactic nuclei (AGNs) over cosmic time
(e.g.,~Ueda \etal\ 2003; Richards \etal\ 2006), but from luminosity
functions alone it is difficult to determine whether mass or
luminosity evolution is the predominant agent of these changes.  One
thing we do know is that the growth of black holes (BHs) and galaxies
are coordinated such that, in local spheroids, BH mass is strongly
correlated with spheroid luminosity (Marconi \& Hunt 2003) and stellar
velocity dispersion (the \msigma\ relation; Gebhardt \etal\ 2000a;
Ferrarese \& Merritt 2000; Tremaine \etal\ 2002; Barth \etal\ 2005).
Charting the mass accretion history of the Universe thus may provide
important insight into the growth of galaxies.  It has become
possible, using the \msigma\ relation, to calibrate virial scaling
relations between AGN luminosity and the size of the broad-line region
(Gebhardt \etal\ 2000b; Ferrarese \etal\ 2001; Onken \etal\ 2004;
Nelson \etal\ 2004; Greene \& Ho 2006b); these techniques have been
used to investigate BH mass functions at intermediate to high redshift
(e.g.,~Vestergaard 2004; McLure \& Dunlop 2004; Kollmeier \etal\ 2006),
but not in a systematic way for nearby systems.  A good measurement of
the local active BH mass function provides an essential boundary
condition for models of the evolution in active BH mass.  Furthermore,
we can probe significantly further down the mass function at the
present day than at any other epoch.

In fact, BH mass functions built from local broad-line AGN samples
have the unique capability to probe BH masses \mbh$\lesssim
10^{6.5}$~\msun.  Mass functions derived from stellar velocity
dispersions (whether they be inactive [Yu \& Tremaine 2002] or
narrow-line active galaxies [Heckman \etal\ 2004]) are necessarily
limited to the current spectral resolution limits of large-area
spectroscopic surveys such as the Sloan Digital Sky Survey (SDSS;
e.g.,~Bernardi \etal\ 2003).  At the same time, the conversion from
galaxy luminosity to BH mass is completely unconstrained at these low
masses.  Direct dynamical mass measurements in this mass range are
beyond the spatial resolving power of current instrumentation for all
but the nearest systems.  Thus broad-line AGNs currently provide the
only means to systematically explore the BH mass function below
$10^6$~\msun\ (``intermediate-mass'' BHs; e.g.,~Greene \& Ho 2004).
Although such objects constitute a negligible fraction of the
present-day BH mass density, it is actually quite important to
characterize the low-mass end of the local BH mass function. For one
thing, it provides one of the only available observational constraints
on models of the initial mass spectrum and halo occupation fraction of
BH seeds in the early Universe (e.g.,~Volonteri \etal\ 2003).
Furthermore, anisotropic gravitational radiation from unequal mass
BH-BH mergers imparts a net linear angular momentum or ``kick'' to the
merger remnant with a velocity that may exceed the escape velocity of
dwarf galaxies (e.g.,~Favata \etal\ 2004; Merritt \etal\ 2004).  We
may test this picture with observational constraints on the number of
local dwarf galaxies that host BHs.

Once we have characterized the zero-redshift broad-line BH mass
function, we may investigate whether broad-line AGNs trace the same
local population as samples selected by alternate means.  For
instance, we expect the soft X-ray luminosity to come from unobscured
sources with broad lines, and thus we expect very similar luminosity
functions for the two populations.  Also, it would be instructive to
compare our results to those of the complementary study of Heckman
\etal\ (2004), which uses the \msigma\ relation to infer the BH mass
function for local narrow-line AGNs.  Our methodology, based on AGN
physics rather than indirectly on the \msigma\ relation, provides an
important alternate measure of the local active BH mass density.
Furthermore, we may compare the space density of narrow- and
broad-line objects as a function of mass; a well-matched comparison
has never been performed in the literature before, and has interesting
consequences for our understanding of AGN unification.  Finally, as we
shall see, the shape of the active and inactive (as inferred from
galaxy luminosity functions) mass functions diverge, both at high and
low mass.  The mass-dependent active fraction contains useful
information about the primary triggering mechanisms of active
galaxies.  By studying in detail the BHs that are radiating in the
present-day Universe, we may hope to gain new insight into the strong
evolution of BH growth with cosmic time.

We briefly review the methodology behind BH mass estimates in 
\S 2, then present our sample selection and methodology in
\S 3, derive the luminosity and mass functions in \S 4, and
discuss the implications in \S 5.  We summarize and conclude in \S 6.
Throughout we assume the following cosmological parameters to calculate
distances: $H_0 = 100~h = 71$~\kms~Mpc$^{-1}$, $\Omega_{\rm m} = 0.27$,
and $\Omega_{\Lambda} = 0.75$ (Spergel \etal\ 2003).

\section{Black Hole Masses For Active Galaxies}

Compared to the dynamical masses used to calibrate the \msigma\
relation, BH mass estimates using AGNs are rather crude, and some
discussion of their validity is in order.  So-called virial masses,
\mbh=$fR\upsilon^2/G$, are estimated using a relation between AGN
luminosity and broad-line region radius ($R$; the radius-luminosity
relation; e.g.,~Kaspi \etal\ 2005), combined with a measurement of the
broad-line region (BLR) velocity dispersion ($\upsilon$).  The factor
$f$ accounts for the unknown broad-line region geometry and here is
assumed to be 0.75, corresponding to a spherical broad-line region
(Netzer 1990).

Although relying on the physics of the BLR to estimate BH masses 
is fraught with danger (e.g.,~Krolik 2001), it is possible to 
directly compare virial masses with the measurements of bulge 
stellar velocity dispersions, and calibrate them with the 
\msigma\ relation.  Rather remarkably, those reverberation-mapped
AGNs with \sigmastar\ measurements show good agreement between 
the two mass indicators (Gebhardt \etal\ 2000b; Ferrarese \etal\ 2001; 
Onken \etal\ 2004; Nelson \etal\ 2004).  Still, the total sample, across 
all above comparisons, consist of no more than 15 objects.  For this 
reason Greene \& Ho (2006b) directly compared \sigmastar\ with 
virial masses for 88 AGNs.  They find the virial masses provide a good 
estimate of BH with a scatter of 0.4 dex, but a number of 
potential systematic effects remain.

The slope of the radius-luminosity relation is currently uncertain,
primarily because it has been measured for only $\sim 30$ objects
spanning a limited luminosity range.  Most recently, Bentz \etal\
(2006) find a shallower radius-luminosity slope ($0.52 \pm 0.04$
rather than the $0.64 \pm 0.02$ assumed here), which would lead to a
narrower distribution in BH mass.  For the typical Eddington ratio of
our sources, \lledd=0.1, the Bentz \etal\ formalism shifts the BH
masses upwards by $\sim 0.3$ dex for a $10^5$~\msun\ BH and downwards
by $\sim 0.2$ dex for a $10^9$~\msun\ BH.  However, as noted by Netzer
\& Trakhtenbrot (2007), if all objects included in the Kaspi \etal\
sample are included in the Bentz \etal\ analysis, the final slope is
consistent with the Kaspi \etal\ result.  Larger reverberation-mapped
samples are urgently needed, but in the meantime we continue to use
the steeper slope.  The other outstanding uncertainty is related to
the (unknown) geometry of the BLR, which translates into a different
pre-factor $f$.  Both Onken \etal\ (2004) and Greene \& Ho (2006b)
find that virial masses calculated assuming a spherical BLR (as we
have done here) are too low by a factor of $\sim 1.6$, corresponding
to a net increase in all BH masses of 0.2 dex.  However, Collin \etal\
(2006) present some evidence that the BLR geometry is luminosity or
Eddington-ratio dependent.  As yet, there are not sufficient
statistics to define a luminosity-dependent $f$ factor, and so we
prefer to apply no correction at this time.

\section{Sample Selection}

The sample selection follows the technique of Greene \& Ho (2004),
which we summarize briefly here.  Our parent sample is drawn from the
Fourth Data Release (DR4) of the SDSS (York \etal\
2002; Adelman-McCarthy \etal\ 2006), which includes spectroscopy of
565,715 objects classified as galaxies and 67,382 objects classified
as AGNs with redshifts $z < 2.1$.  We analyze
all spectra with $z <0.352$, although we remove those galaxies for
which $>20\%$ of the pixels between 6400--6700 \AA\ have been flagged
as bad by the SDSS pipeline; this results in a total of 544,127
spectra (including multiple observations of the same object).
Following Greene \& Ho (2004; see also Ho \etal\ 1997a), we begin by
modeling and removing the stellar continuum, which may mask or mimic
the presence of broad emission lines that we use as the signature of
an accreting BH.  We use the principal component analysis (PCA) method
of Hao \etal\ (2005a) that encapsulates the variance in SDSS
absorption-line galaxies in a set of eight orthogonal eigenspectra.
The galaxy continua are modeled as linear combinations of these
eigenspectra, a power-law to represent the AGN continuum, and a
possible A-star component.  We have not attempted to model the broad
\feii\ emission that is present throughout most of the optical
spectrum (e.g.,~Boroson \& Green 1992; Greene \& Ho 2006a); while this
is important for measurements of \hbeta\ and \oiii, the impact on the
\halpha\ region is negligible.

After galaxy subtraction, in order to limit the number of potential
candidates, we employ an algorithm that identifies potential broad
emission in the \halpha\ region, based on an excess variance in the
\halpha\ region compared with the featureless continuum.  The
following analysis is performed over the spectral region 6400--6700
\AA.  Narrow emission-line flux is removed in an iterative scheme, in
which points with fluxes $> 4~\sigma$ (as measured in the entire
spectral region) are replaced with the median value in the region,
until there is $< 1\%$ decrease in $\sigma$, in practice taking no
more than 5 iterations.  This method of narrow emission line removal
is conservative, in the sense that some flux from very strong narrow
emission lines will remain.  However, without a full spectral
decomposition of the region, which we perform below, a single-Gaussian
fit to the narrow emission lines may easily remove a relatively narrow
broad component.
We then smooth the spectrum with a 10-pixel boxcar
to remove high-frequency noise and increase our sensitivity to real
broad components.  Those sources with a root-mean-square (rms)
deviation that is 50\% higher in the 6400--6700 \AA\ region than in
the regions 5500--6280\AA\ (excluding 20 \AA\ around the NaD $\lambda
5895$ doublet) and 6900--7300 \AA\ are kept, for a total of 31,750
spectra.

%%%%%%%%%%%%%%%%%%%%%%%%%%%%%%%%%%%%%%%%%%%%%%%%%%%%%%%%%%%%%%%%%%%%
%%BoundingBox: 
\begin{figure*}
\vbox{ 
\vskip -0.1truein
\hskip 0.in
\psfig{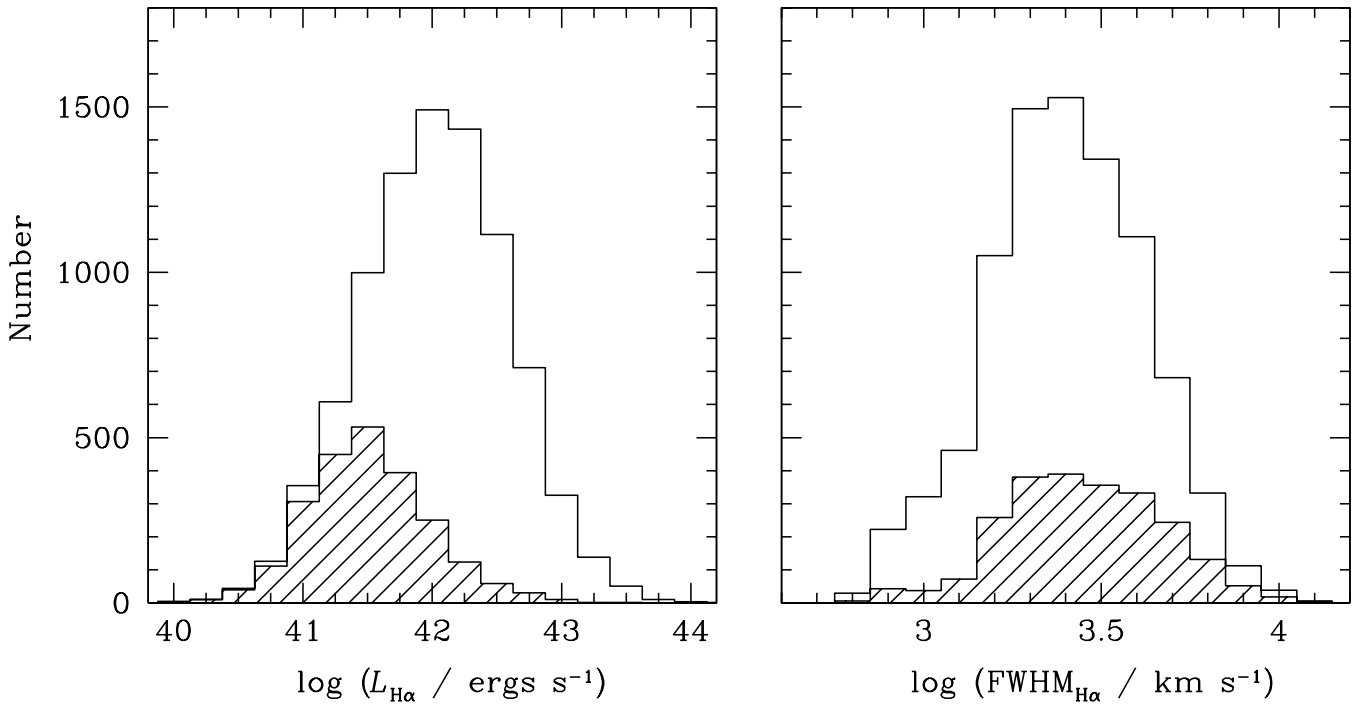}}
\vskip -0mm
\figcaption[]{
Distributions of broad \lha\ ({\it left}) and broad \fwha\ ({\it
right}) for the entire sample of broad-line AGNs.  The shaded
histograms indicate the distribution of those objects targeted within
the SDSS main galaxy sample.  Note the dominance of objects with
\fwha$<2000$~\kms, the canonical division between classical and
``narrow-line'' Seyfert 1 galaxies.
\label{masshist}}
\end{figure*}
%\vskip 5mm
%%%%%%%%%%%%%%%%%%%%%%%%%%%%%%%%%%%%%%%%%%%%%%%%%%%%%%%%%%%%%%%%%%%%%

With this smaller subset of sources we perform more detailed modeling
of the \halpha+\nii~$\lambda \lambda 6548,~6583$ region.  Careful
modeling of the narrow lines is crucial, since velocity structure in
the narrow-line region can mimic broad lines (see additional arguments
in Ho \etal\ 1997c and details of the methodology in Greene \& Ho
2004, 2005a,b).  Whenever possible, we build a multi-component
Gaussian model for the narrow lines using the nearby \sii~$\lambda
\lambda 6716,~6731$ lines that empirically are found to provide a good
model for the narrow \hn\ (Filippenko \& Sargent 1988; Ho \etal\
1997c), and in cases without \sii\ emission, we resort to using a
model based on the core of the \oiii~$\lambda 5007$ line.  The
relative wavelengths of \halpha\ and the \nii\ doublet are fixed to
laboratory values, as are the relative strength of the \nii\ lines,
and only the relative amplitudes of the \halpha\ and \nii\ lines are
allowed to vary. We adopt the empirical criterion of Hao \etal\
(2005a) to determine whether an additional component is required: any
component that results in a $20\%$ decrease in \chisq\ is deemed
statistically justified.  Broad lines are often lumpy and asymmetric,
so we model the broad emission with multiple Gaussians, but attach no
physical significance to any given component.  All objects with broad
\halpha\ components apart from those in the narrow-line model are
retained.

%%%%%%%%%%%%%%%%%%%%%%%%%%%%%%%%%%%%%%%%%%%%%%%%%%%%%%%%%%%%%%%%%%%%
%%BoundingBox: 
\hskip -0.15in
\psfig{file=sample.epsi,width=0.5\textwidth,keepaspectratio=true,angle=0}
\vskip 4mm
%%%%%%%%%%%%%%%%%%%%%%%%%%%%%%%%%%%%%%%%%%%%%%%%%%%%%%%%%%%%%%%%%%%%%
%\noindent

%%%%%%%%%%%%%%%%%%%%%%%%%%%%%%%%%%%%%%%%%%%%%%%%%%%%%%%%%%%%%%%%%%%%
%%BoundingBox: 
\begin{figure*}
\vbox{ 
\vskip -0.2truein
\hskip 0.in
\psfig{file=mass_ledd.epsi,width=0.5\textwidth,keepaspectratio=true,angle=-90}}
\vskip -0mm
\figcaption[]{
Distributions of BH mass ({\it left}) and \lledd\ ({\it right}) for
the entire sample of broad-line AGNs.  \mbh\ is calculated from
\lha\ and \fwha\ as shown above, using the formalism of Greene \& Ho
(2005b), and the Eddington ratio is derived assuming an average
bolometric correction of $L_{\rm{bol}} = 2.34 \times 10^{44} (L_{\rm
H\alpha} / 10^{42})^{0.86}$ \lum\ (see text).  As in the previous figure,
the galaxies targeted in the SDSS main sample are indicated with the
shaded histogram.  Note that while the galaxy-selected AGNs are
uniformly at low luminosities, they span the entire range of \mbh\ in
the sample.  Massive BHs are predominantly in low-accretion states at
the present time.
\label{fwhm}}
\end{figure*}
%\vskip -5mm
%%%%%%%%%%%%%%%%%%%%%%%%%%%%%%%%%%%%%%%%%%%%%%%%%%%%%%%%%%%%%%%%%%%%%

Our goal is to limit our sample to those targets for whom we may
estimate a reliable BH mass from broad \halpha.  We thus very
aggressively flag all suspicious objects based on their line strength,
by applying a combined cut on total \halpha\ flux normalized to the
rms deviations in the continuum-subtracted spectrum, and the \halpha\
equivalent width (EW).  We require that \fha/rms $> 200$ and
EW(\halpha) $>$ 15 \AA.  The EW measurement is based on continuum
measured from the best-fit PCA galaxy model in the spectral regions
6570--6590 \AA\ and 6480--6540 \AA.  In detail, our thresholds are
selected based on simulations, presented in Appendix A, which
demonstrate that for the S/N ratios typical in the SDSS, BH mass
estimates are highly uncertain ($>1$ dex) and significantly biased for
lower \halpha\ fluxes and EWs.  Hereafter, this combination of cuts
will be referred to as the ``detection threshold.''  Following the
automated flagging, some objects at the low-mass end are still
questionable when examined by eye.  They are flagged as well.  A total
of 11,428 objects are removed at this stage.  For comparison with
previous work, we also build a more inclusive luminosity function that
does not reject the flagged objects.

There are a number of scientifically interesting objects that fall
below the detection threshold.  There are strongly star-forming
galaxies with a broad \halpha\ base (see also Hao \etal\ 2005a) that
are probably Wolf-Rayet galaxies (Ho \etal\ 1997c).  In other cases,
we are probably seeing a small, scattered component of broad emission
in what is predominantly a narrow-line AGN.  In this case, the
\halpha\ luminosity does not correlate with the broad-line region
size.  Finally, if the BH is radiating at a very low fraction of its
Eddington luminosity then it is thought that the standard optically
thick, geometrically thin accretion disk (Shakura \& Sunyaev 1973) is
replaced by a radiatively inefficient accretion mode (e.g.,~Quataert
2001; Narayan 2006 and references therein).  Although no such objects
are included in the reverberation-mapped samples used to calibrate the
radius-luminosity relation (e.g.,~Kaspi \etal\ 2005) it is likely that
the broad-line region structure is quite different in this
regime. While we do not have sufficient S/N in single objects to
accurately measure BH masses, it is quite possible that in stacked
spectra we would be able to uncover very low-contrast broad lines.  
We defer such analysis to a later work.

Finally, we note that narrow-line AGN samples are typically selected
using their location in diagnostic diagrams (Baldwin \etal\ 1981;
Veilleux \& Osterbrock 1987; Ho \etal\ 1997a; Kauffmann \etal\ 2003;
Hao \etal\ 2005a) that discriminate the shape of the ionizing
continuum by the relative strengths of various prominent narrow
emission lines.  Our selection technique relies on the presence of
broad emission lines, and so is unbiased with respect to line ratios.
This is particularly useful for selecting relatively metal-poor AGN
hosts, since the typically high ratio of \nii/\halpha\ may be
significantly reduced.  This has been observed in NGC 4395
(e.g.,~Kraemer \etal\ 1999), with low significance in the original
Greene \& Ho (2004) sample, and quite strikingly in our new sample of
Type 2 low-mass BHs (A.~J.~Barth \etal, in preparation; see also
Groves \etal\ 2006).  In contrast to standard narrow-line selections,
therefore, our selection is relatively unbiased against AGNs in
low-metallicity (and thus typically low-mass; e.g.,~Tremonti \etal\
2004) galaxies in this manner.

Our final sample (with duplicates removed) of broad-line AGNs is
composed of 8728 objects.  The Princeton spectral reductions include
single-Gaussian fits to narrow emission lines, and flags for all
objects with an additional broad component.  Approximately $30\%$ of
the broad-line objects in their sample are rejected by our algorithms,
some ($35\%$) with the initial \halpha-finding algorithm, and the rest
either because of a lack of broad \halpha\ in the multi-Gaussian fit,
or because they fall below our detection threshold.  We have examined
a large number of these objects visually, and we feel justified in
rejecting them.  One contaminant in particular consists of objects
with multi-component narrow lines, which confuse algorithms that fit
only single Gaussians to the narrow lines.  
We also compare with the
Hao \etal\ (2005a) list of broad-line AGNs.  Only $53\%$ of their
broad-line sample is included in our final sample; $17\%$ are rejected
by our initial broad-line search algorithm, $11\%$ are deemed to have
no broad \halpha\ in more detailed fitting, and $19\%$ are rejected by
our detection threshold.

We have measured the \halpha\ luminosities (with no internal reddening
correction applied) and \fwha\ for our entire sample of broad-line
AGNs (Fig. 1).  The \fwha\ are measured from the multi-component
Gaussian models (as in Greene \& Ho 2004).  We can use these measured
quantities and the formalism presented in Greene \& Ho (2005b) to
derive BH masses for the sample.  We also use the measured \halpha\
luminosities and the inferred BH masses to estimate Eddington ratios
for the sample, \lledd, where $L_{\mathrm{Edd}} \equiv 1.26 \times
10^{38}$~(\mbh/\msun) \lum. Assuming that $L_{\rm{bol}} = 9.8$ \lf\
(McLure \& Dunlop 2004), in terms of \lha\ our bolometric correction
is $L_{\rm{bol}} = 2.34 \times 10^{44} (L_{\rm H\alpha} /
10^{42})^{0.86}$ \lum\ (Greene \& Ho 2005b).  We adopt this bolometric
correction throughout.  The corresponding distribution of BH masses
and Eddington ratios is shown in Figure 2.  At face value the mass
function is strongly peaked at $10^7$~\msun, while the typical
Eddington ratio of our sample is about a tenth of the Eddington limit.
However, there are severe selection effects that cause us to lose
significant numbers of sources, and these depend on mass, Eddington
ratio, and redshift.  The strong importance of selection effects is
shown in Figure 3, where the distribution of BH mass in the sample is
shown as a function of both redshift and Eddington ratio.  The most
striking trends in this figure are the decrease of characteristic BH
mass as one moves to higher Eddington
ratio at a given redshift, and the increase of characteristic mass as
one moves to higher redshift at a given Eddington ratio.  Both of
these trends are driven by the magnitude limits of the SDSS.

%%%%%%%%%%%%%%%%%%%%%%%%%%%%%%%%%%%%%%%%%%%%%%%%%%%%%%%%%%%%%%%%%%%%
%%BoundingBox: 
%\begin{figure*}
%\vbox{ 
%\vskip -0.3truein
\hskip -0.2in
\psfig{file=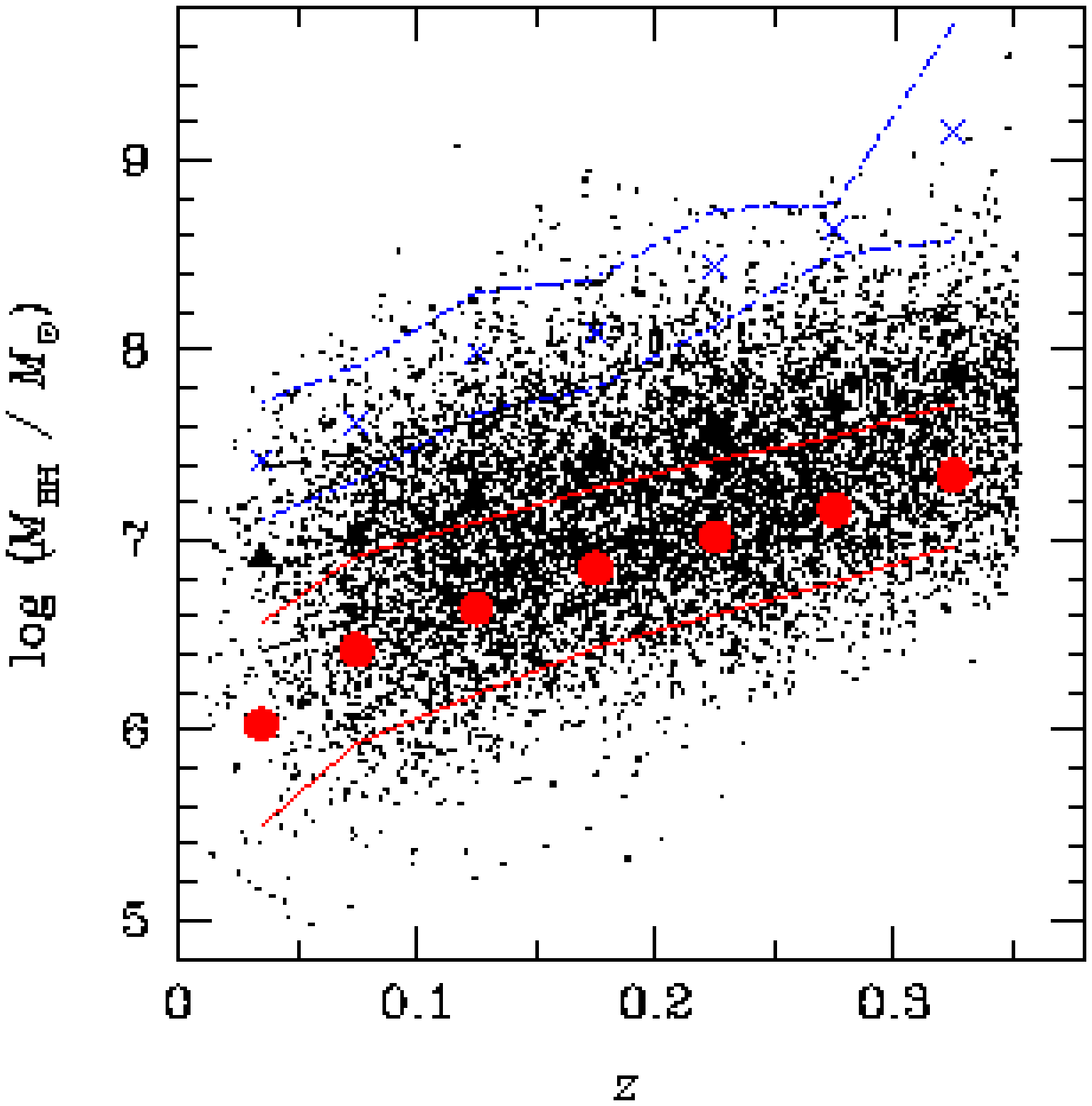,width=0.47\textwidth,keepaspectratio=true,angle=0}
%}
\vskip -1mm
\figcaption[]{Distribution of \mbh\ with redshift is shown in small points 
for the entire sample.  In large symbols we represent the mean mass in 
three different Eddington ratio bins, with $1~\sigma$ contors bracketing 
each bin.  The blue crosses and dash-dotted lines show objects 
with \lledd$<0.01$, the black triangles and dashed lines show
$0.01<$\lledd$<0.1$, and the red circles and solid lines are objects 
with \lledd$>0.1$.}  
%\end{figure*}
\vskip 4mm
%%%%%%%%%%%%%%%%%%%%%%%%%%%%%%%%%%%%%%%%%%%%%%%%%%%%%%%%%%%%%%%%%%%%%
%\noindent

%%%%%%%%%%%%%%%%%%%%%%%%%%%%%%%%%%%%%%%%%%%%%%%%%%%%%%%%%%%%%%%%%%%%
%%BoundingBox: 
%\begin{figure*}
%\vbox{ 
%\vskip -0.4truein
\hskip -8mm
\psfig{file=lumfunc_all_v6mod.epsi,width=0.5\textwidth,keepaspectratio=true,angle=0}
%}
%\vskip -0mm
\figcaption[]{
Volume-weighted broad \halpha\ luminosity function in bins of 0.25 dex
(\# Mpc$^{-3}$ log \lha$^{-1}$).  The maximum volume is calculated
based on both the photometric and spectroscopic limits of the
survey and our search algorithm (see text).  The error-bars represent
the Poisson errors in each bin.  The solid line represents our
maximally inclusive sample, in which objects with less reliable broad
\halpha\ are included.  Our best-fit double power law is shown as the
dashed line.  The inset panels show the luminosity functions for
objects targeted as galaxies or AGNs, respectively, primarily based on
a color selection.
\label{lumfunc}}
%\end{figure*}
%%%%%%%%%%%%%%%%%%%%%%%%%%%%%%%%%%%%%%%%%%%%%%%%%%%%%%%%%%%%%%%%%%%%
\vskip 5mm
%\noindent

\section{Luminosity and Mass Functions}

\subsection{Broad \halpha\ Luminosity Function}

We begin by calculating a traditional luminosity function $\Phi$(\lha)
for our broad-line active galaxy sample.  While the luminosity
function is difficult to interpret directly, in so far as it is a
convolution of the BH mass and Eddington ratio distributions,
nevertheless it has the benefit of being a directly observable
quantity.  BH masses, on the other hand, are potentially subject to
considerable systematic uncertainties (e.g.,~Greene \& Ho 2006b;
Collin \etal\ 2006).  Furthermore, the luminosity function is readily
compared with various probes 
of the active galaxy population, such as the X-ray luminosity function
for AGNs.  Following common practice, we present the differential
luminosity function, the number of AGNs per unit volume per
logarithmic luminosity interval, $\hat{\Phi}$(\lha) $=($\lha$/{\rm
log}_{10}e)\Phi$(\lha).  Since galaxies of different magnitude are
observable to varying distance within the sample, a proper
representation of the distribution of AGN luminosity (or mass)
requires that we somehow account for our variable sensitivity prior to
comparing their distributions directly.  We primarily use the
classical \vvmax\ weighting method (Schmidt 1968; Huchra \& Sargent
1973; Condon 1989; Ulvestad \& Ho 2001), but in Appendix B we present
a non-parametric maximum likelihood luminosity function as well
(e.g.,~Efstathiou \etal\ 1988).

%%%%%%%%%%%%%%%%%%%%%%%%%%%%%%%%%%%%%%%%%%%%%%%%%%%%%%%%%%%%%%%%%%%%%
%%%%%%%%%%%%%%%%%%%%%%%%%%%%%%%%%%%%%%%%%%%%%%%%%%%%%%%%%%%%%%%%%%%%
%%BoundingBox: 
%\begin{figure*}
%\vbox{ 
%\vskip -0.4truein
\hskip -8mm
\psfig{file=lumfunc_maxlik_v7.epsi,width=0.5\textwidth,keepaspectratio=true,angle=0}
%}
%\vskip -0mm
\figcaption[]{
Broad \halpha\ luminosity function calculated using the maximum
likelihood formalism (solid line; see Appendix B for details).  The
\vvmax\ luminosity function is shown in solid symbols as in Figure 4.
Also shown are the single (dashed line) and double (dot-dashed line)
power-law fits to the Hao \etal\ (2005b) maximum likelihood total
\halpha\ luminosity function.
\label{lumfunc}}
%\end{figure*}
\vskip 5mm
%%%%%%%%%%%%%%%%%%%%%%%%%%%%%%%%%%%%%%%%%%%%%%%%%%%%%%%%%%%%%%%%%%%%%
%\noindent

In the \vvmax\ formalism, our sensitivity is characterized as the
maximum volume to which each source would be included in our final
sample.  In calculating \vmax, we must account for both the magnitude
limit of the SDSS and the S/N dependence of our \halpha\ detection
procedure.  The appropriate magnitude limit will depend on whether a
given object was targeted as a galaxy or quasi-stellar object (QSO).
The main galaxy sample comprises spatially resolved (i.e.,
non-stellar) targets above a limiting $r$-band Petrosian magnitude of
17.77 mag (Strauss \etal\ 2002).  QSO targets, in contrast, are
selected with a rather complicated set of color-selection criteria
within a magnitude range of $15.0 < i < 19.1$ (Richards \etal\ 2002;
Stoughton \etal\ 2002).  The maximum flux limit mitigates bleeding
between fibers.  A different color criterion is used to select
high-redshift QSO candidates, and these have a limiting magnitude of
$i = 20.2$; a small number (237) of our final targets fall into this
category.  We note that it is 
possible for objects to have been targeted as either a QSO or a
galaxy; these objects are considered part of the galaxy sample.  Now,
in addition to these two primary surveys, some ``serendipitous''
sources are targeted, due to a detection either in the \rosat\ All-Sky
Survey (Voges \etal\ 1999), or by the FIRST radio survey (Becker
\etal\ 1995; see also Richards \etal\ 2002).  We have 154 such
objects; they do not constitute a complete sample, nor do they have a
uniform detection limit.  We therefore assign them a maximum volume
corresponding to their observed volume (i.e., take $V_{\rm max}=V$),
and then verify that our conclusions are unchanged when we exclude
them from the sample.

%%%%%%%%%%%%%%%%%%%%%%%%%%%%%%%%%%%%%%%%%%%%%%%%%%%%%%%%%%%%%%%%%%%%
%%BoundingBox: 
%\hskip 0.in
\psfig{file=table_lumfunc.epsi,width=0.4\textwidth,keepaspectratio=true}
\vskip 5mm
%%%%%%%%%%%%%%%%%%%%%%%%%%%%%%%%%%%%%%%%%%%%%%%%%%%%%%%%%%%%%%%%%%%%%
%\noindent

Based on the target selection of each object, it is straightforward to
calculate the maximum volume to which the SDSS would have
spectroscopically targeted a given object, with approximate
k-corrections based on the observed colors.  In detail, of course, the
color (and thus target selection) of a source is redshift-dependent,
as the spatially resolved galaxy light will dim more rapidly than the
unresolved AGN light.  This is difficult to model without a good
separation of host and AGN light.  Nevertheless, since the majority of
the galaxy targets are very close to their limiting magnitude, this
should not affect the results in practice.

With increasing redshift and commensurately degraded S/N, it becomes
more and more challenging to detect broad \halpha.  At a certain
point, the object reaches our imposed detectability threshold, and is
removed from the sample.  In detail, however, our algorithm may lose
the ability to detect the object even before this limit is 
reached.  This depends on the contrast of the broad line, the galaxy
subtraction, the strength of the narrow lines, etc.  In principle, it
is also possible that the galaxy dilution increases as the physical
radius subtended by the fixed SDSS fiber increases, although it will
be mitigated to some degree by surface brightness dimming, and we have
neglected this effect.  To model our incompleteness, we generate
artificial spectra over a grid of redshifts (S/N) between the observed
redshift and either the photometric redshift limit or the
detectability threshold redshift, whichever is smaller.  The S/N at
each interval is calculated based on the fiber magnitude of the source
at the new redshift and an empirical relation between fiber magnitude
and S/N we have derived from the SDSS data.  If the interval is
smaller than $\delta z = 0.05$, we simply bisect the redshift
interval, whereas for larger redshift ranges we generate artificial
spectra at $z_{\rm max}-0.05$ and the center of the redshift interval.
We generate three spectra at each redshift, with a range of S/N $\pm
20\%$ of the calculated value.  When the calculated S/N is $\leq$ 5,
the limiting redshift is automatically reached, as such spectra are
simply unusable.  We subject the remaining spectra to our full
selection algorithm.  As long as two of the three artificial spectra
are recognized as broad-line AGNs, we continue to the next redshift
bin.  In practice, only $\sim 20\%$ of our \vmax\ calculations are
lowered by these simulations, suggesting that our detection threshold
is reasonable.

Using the \vvmax\ weights derived above, the luminosity function 
($\hat{\Phi}$) and uncertainty per bin ($\sigma$) are simply
calculated:
\begin{equation}
\hat\Phi(L_{\rm H\alpha}) = 
\sum_i^{N}\left( \frac{1}{V_{\rm max}} \right)_i, \\
\sigma = \left[\sum_i^{N}\left( \frac{1}{V_{\rm max}} 
\right)^2_i \right]^{1/2};
\end{equation}
they are shown in Table 2 and Figure 4.  
The survey volume is determined assuming an angular coverage of 4783
deg$^2$ of the DR4 spectroscopic survey\footnote{{\tt
http://www.sdss.org/dr4/}}. We compute the number of AGNs per unit
volume, per unit logarithmic luminosity [$\hat{\Phi}$(\lha)] in bins
of 0.25 dex in \halpha\ luminosity, from $10^{40}-10^{43.25}$~\lum.
In the insets are plotted the luminosity functions for objects
targeted as galaxies ({\it left}) and QSOs ({\it right}).  The
``galaxies'' are strongly peaked at the luminosity of a $10^7$~\msun\
BH radiating at 10\% of Eddington.  At higher luminosity, the QSOs
tend to dominate the light, while at the lowest luminosities, there
are equal numbers of each type, depending on whether the source is a
low-mass BH at higher Eddington ratio (QSO) or a more massive and less
active source.  Note that the decline at luminosities below $\lesssim
10^{40.75}$ ergs s$^{-1}$ is due to incompleteness.  For instance, the
Palomar spectroscopic survey of nearby galaxies, with significantly
higher sensitivity to low-level broad emission, finds a luminosity
function that continues to rise to significantly lower luminosities
(Ulvestad \& Ho 2001; Ho 2004).  For this reason, we also calculate a
maximally inclusive luminosity function, retaining all flagged
objects.  That is shown as a solid line in Figure 4.

%%%%%%%%%%%%%%%%%%%%%%%%%%%%%%%%%%%%%%%%%%%%%%%%%%%%%%%%%%%%%%%%%%%%
%%BoundingBox: 
%\begin{figure}
%\vbox{ 
\vskip -0.05truein
\hskip -4mm
\psfig{file=lumfunc_xray.epsi,width=0.45\textwidth,keepaspectratio=true,angle=0}
%}
\vskip -0mm
\figcaption[]{
Conversion of our observed luminosity function to an X-ray luminosity function
using the Strateva \etal\ (2005) relation between UV luminosity and \aox.  
Solid line is the soft X-ray, unabsorbed luminosity function presented in
Hasinger \etal\ (2005).  The points represent the median luminosity
function of 1000 Monte Carlo realizations of the optical to X-ray
conversions, which account for its significant scatter.  The dotted
lines bracket 68\% of the resulting distribution in a given luminosity
bin.}
%\label{fwhm}}
%\end{figure}
\vskip 5mm
%%%%%%%%%%%%%%%%%%%%%%%%%%%%%%%%%%%%%%%%%%%%%%%%%%%%%%%%%%%%%%%%%%%%%
%\noindent

For comparison with other samples it is useful to parameterize the
observed luminosity function.  Following previous work, we
consider a Schechter (1976) function and a double power law
(e.g.,~Croom \etal\ 2004; Hao \etal\ 2005b),
\begin{equation}
\phi (L) = \frac {\phi^*(L_*) / L_*}{(L/L_*)^{\alpha}+(L/L_*)^{\beta}},
\end{equation}
where $\phi^*(L_*)$, $L_*$, $\alpha$ and $\beta$ are free parameters.
We use a \chisq\ minimization fitting procedure that accounts for the
asymmetric errors in space density, and we fit all objects in our
final sample with luminosities $\geq 10^{40}$~\lum\ (for this
luminosity range the results do not change if we fit the inclusive
sample).  The Schechter fit has a reduced \chisqr$=11$, while the
double power-law fit has a reduced \chisqr$=1.3$.  Therefore, we
prefer the latter, which yields values of $\phi (L_*) = (2.0 \pm 0.4)
\times 10^{-6}$ Mpc$^{-3}$, $L_* = 10^{42.0 \pm 0.1}$~\lum, $\alpha
= 1.29 \pm 0.09$, and $\beta = 2.82 \pm 0.07$.  This fit is shown as a
dashed line in Figure 4.

\subsection{Comparison with Hao \etal\ (2005b)}

Hao \etal\ (2005b) used a smaller subsample of the SDSS to compute
emission-line luminosity functions for broad- and narrow-line AGNs.
They used very comparable selection criteria to select their
broad-line subsample, and thus it is important to compare our results
with theirs.  In order to compare directly with their work we
construct an additional luminosity function using the
maximum likelihood formalism (for details, see Appendix B).  At high
luminosity, Hao \etal\ have limited statistics (as demonstrated by the
divergence of their single and double power-law fits in Fig. 5), but
the double power-law fit does provide a better fit to their highest
luminosity bins, consistent with our result.  At lower luminosity, our
function diverges from that of Hao et al.  We believe this discrepancy
may be attributed to three factors.  

Firstly, we consider flux only from broad \halpha, while Hao \etal\
include both narrow and broad emission.  This has no effect at high
luminosity, where the narrow-line luminosity constitutes a negligible
fraction of the total, but at lower luminosities, objects with high-EW
narrow lines will be preferentially shifted into higher luminosity
bins in the Hao \etal\ function.  Secondly, we remove a large number
of broad-line AGNs below our detection threshold.  However, even when
we include these objects, our luminosity function flattens at
considerably higher luminosity than the Hao \etal\ function.  Relative
to Hao \etal\, we have removed objects both with our initial \halpha\
detection algorithm and with our line-fitting procedure, at the $\sim
20 \%$ level.  Because the luminosities involved are very low, the
corresponding volume corrections are very large, leading to highly
discrepant space densities in the two studies.  In some respects it is
not surprising to find such a large range of possible space densities
at these very low \halpha\ luminosities.  As we show in Appendix A,
the measurements of \lha\ are very inaccurate in this regime.  Perhaps
the most important message of this exercise is that the SDSS is not an
adequate tool to measure the line luminosity function below $\sim
10^{40.5}$~\lum.

In addition to the differences in object selection, we adopt a very
different set of assumptions in building our luminosity functions than
Hao et al.  They have used the maximum likelihood formalism, but in
calculating their selection function [$p_i(L)$] have assumed that
neither the line shape nor the galaxy luminosity are linked to the \halpha\
luminosity.  In our \vvmax\ formalism, we need not make any assumption
about galaxy luminosity.  However, when we calculate our selection
function for the maximum likelihood function, we do not allow a given
object to exceed its Eddington luminosity, thus implicitly
assuming a correlation between galaxy and AGN luminosity.  For the
majority of objects, we are thus adopting a higher maximum luminosity
than the fiducial luminosity of $\sim 10^{42.6}$~\lum\ assumed by Hao \etal.
Furthermore, we assume that as the AGN luminosity decreases for a
given object, the line width increases, thus increasing the noise per
pixel across the broad line, and decreasing our detection efficiency
at low luminosity.  The net effect at low luminosity is apparently
that we have larger volume corrections than Hao \etal, leading to
lower inferred space densities.  

As an aside, we note that our luminosity function is consistent with
that of Croom \etal\ (2004), with the break luminosity $\sim 1$ mag
lower.  On the other hand, while the Hao \etal\ function appears to be
consistent with that of Ulvestad \& Ho (2001; see Fig. 11 of Hao
\etal), the galaxy luminosity is included in the latter luminosity
function.  In most cases the AGN luminosity itself is significantly
fainter, and it is probably fair to shift their function fainter by
$\sim 4$ magnitudes (see Ho 2004 for details).

\subsection{X-ray Luminosity Function}

As an additional sanity check, we compare our luminosity function with
the zero-redshift luminosity functions presented by Hasinger \etal\
(2005).  This work focuses predominantly on soft X-ray--selected
samples with broad Balmer lines in optical follow-up (see also Schmidt
\etal\ 1998).  To perform a comparison, we (statistically) convert our
observed \halpha\ luminosities to soft X-ray luminosities.  We employ
the Greene \& Ho (2005b) conversion between \halpha\ luminosity and
\lf, and then, assuming a spectral slope of $\alpha = 0.44$ ($f_{\nu}
\propto \nu^{-\alpha}$; see, e.g., Vanden Berk \etal\ 2001; Greene \&
Ho 2005b), the relation between \aox\ and \luv\ from Strateva \etal\
(2005).  Although the Strateva \etal\ relation (see also Steffen
\etal\ 2006) is only measured for UV luminosities of $\sim
10^{43}-10^{48}$~\lum, Greene \& Ho (2007) find that, in general, the
low-mass AGNs from Greene \& Ho (2004) obey the extrapolation of the
Strateva relation to lower luminosity.  Because there is large scatter
in the conversion between X-ray and optical luminosity, we generate
1000 X-ray luminosity functions from our optical function, in each
case perturbing the intercept and slope of the relation by a
log-normal deviate determined by the measured scatter in each
variable.  We then select the median value at each luminosity, and the
upper and lower bounds to bracket 68\%.  The resulting inferred soft
X-ray luminosity function is shown in Figure 6, with the Hasinger
\etal\ double power-law fit shown as a solid line.  Given the
uncertainties involved, we find quite reasonable agreement.

%%%%%%%%%%%%%%%%%%%%%%%%%%%%%%%%%%%%%%%%%%%%%%%%%%%%%%%%%%%%%%%%%%%%
%%BoundingBox: 
\hskip 0.1in
\psfig{file=table_massfunc.epsi,width=0.3\textwidth,keepaspectratio=true}
\vskip 4mm
%%%%%%%%%%%%%%%%%%%%%%%%%%%%%%%%%%%%%%%%%%%%%%%%%%%%%%%%%%%%%%%%%%%%%
%\noindent

\subsection{BH Mass Function}

Using the \vmax\ weights derived above, we show in Figure 7
volume-weighted BH mass function (the number of BHs per unit volume,
per unit logarithmic BH mass, in bins of 0.25 dex in BH mass; see also
Table 3).  As with the luminosity function, we fit the mass function
with a Schechter function, a double power law, and a log-normal
parameterization.  In this case we include all mass bins in the
fitting.  Once again, the Schechter function provides a poor fit, with
\chisqr~$=18$.  Both the double power-law and log-normal functions
provide reasonable fits, with \chisqr~$=2.5$ and \chisqr~$=1.2$,
respectively.  The best-fit double power law is shown as a dashed line
in Figure 7, and is parameterized by $\phi (M_*) = (4.7 \pm 0.5)
\times 10^{-6}$ Mpc$^{-3}$, $M_* = 10^{7.32 \pm 0.05}$~\msun, $\alpha
= 0.78 \pm 0.06$, and $\beta = 3.00 \pm 0.07$.  The log-normal fit,
shown as a solid line in Figure 7, has a normalization of $(6.4 \pm
0.3) \times 10^{-6}$ Mpc$^{-3}$, a central mass of $10^{6.6 \pm
0.03}$~\msun, and a width of $0.59 \pm 0.01$ dex.

Taken at face value, our BH mass function displays a clear break at
$\sim 10^{6.6}$~\msun, and declines toward both higher and lower mass.
This is comparable to the characteristic mass of local narrow-line
AGNs (Heckman \etal\ 2004) but is significantly lower than that of
inactive BHs ($\sim 10^8$~\msun; e.g.,~Marconi \etal\ 2004).
Unfortunately, as demonstrated by Figure 3, our data are dominated by
the strong selection bias inherent in a magnitude-limited sample.  It
is clear that, in terms of luminosity, we are significantly incomplete
below $\sim 10^{40.75}$~\lum.  However, incompleteness in luminosity
is redshift-dependent and may not be translated directly into
incompleteness as a function of BH mass.  At a given \lha, the S/N
across the line (and thus its detectability) decreases with increasing
\fwha, and thus increasing BH mass.  Additionally, as the BH mass
increases, the corresponding bulge luminosity is presumably
increasing, further decreasing the contrast and the S/N in the broad
line.

%%%%%%%%%%%%%%%%%%%%%%%%%%%%%%%%%%%%%%%%%%%%%%%%%%%%%%%%%%%%%%%%%%%%
%%BoundingBox: 
%\begin{figure*}
%\vbox{ 
%\vskip -0.3truein
\hskip -0.2in
\psfig{file=massfunc_all_vmaxbest_v6mod.epsi,width=0.5\textwidth,keepaspectratio=true,angle=0}
%}
%\vskip -0mm
\figcaption[]{
Volume-weighted BH mass function in bins of 0.25 dex (\# Mpc$^{-3}$
log \mbh$^{-1}$).  The weights used are identical to those for the
luminosity function, and as above we show in the inset the mass
functions for objects targeted as galaxies ({\it left}) and QSOs ({\it
right}), respectively.  Although we are subject to significant
incompleteness, we will argue below that there is truly a turnover in
active galaxy masses at both lower and higher BH masses.  We have fit
the mass function with both a double power-law (dashed line) and a
log-normal function (solid line).
\label{bhmfunc}}
%\end{figure*}
\vskip 2mm
%%%%%%%%%%%%%%%%%%%%%%%%%%%%%%%%%%%%%%%%%%%%%%%%%%%%%%%%%%%%%%%%%%%%%
%\noindent

If we could uniquely ascribe a host galaxy luminosity and light
profile to each \mbh, then we could easily model our incompleteness as
a function of \mbh\ and \lledd.  At high BH mass, this is in fact
possible, as there is a relation linking \mbh\ and spheroid
luminosity, and the fundamental plane tells us the typical sizes (and
thus fiber luminosities) of elliptical galaxies.  However, for spiral
or dwarf spheroidal host galaxies, there ceases to be a unique mapping
between \mbh\ and galaxy luminosity or structure.  The relation
between bulge-to-total ratio and galaxy luminosity is poorly
quantified and contains significant scatter in any case.
Furthermore, at lower masses, as the AGNs become intrinsically fainter,
only systems with relatively luminous host galaxies will fall above
the magnitude limit of the SDSS.  For these reasons, at low \mbh\ the
calculated incompleteness is a strong function of the assumed (but
unconstrained) host galaxy morphology.

%%%%%%%%%%%%%%%%%%%%%%%%%%%%%%%%%%%%%%%%%%%%%%%%%%%%%%%%%%%%%%%%%%%%
%%BoundingBox: 
\begin{figure*}
\vbox{ 
\vskip -0.1truein
\hskip -0.2in
\psfig{file=compl_hist_low.epsi,width=0.35\textwidth,keepaspectratio=true,angle=-90}
}
\vskip -0mm
\figcaption[]{
Incompleteness as a function of galaxy luminosity for simulated galaxies.
These simulations are run for BH masses of \mbh$=10^{5.5}-10^{6.4}$~\msun\ 
at three redshifts ($z=0.04,~0.07,~0.10$), and with the limited range of   
\lha$=10^{40.5}-10^{41}$~\lum.  Although the input fiber luminosities 
are $-14 \leq M_{B} \leq -22$, there are non-zero detection fractions 
only over the range of fiber luminosities shown here.  The simulations 
demonstrate that we have roughly constant incompleteness 
over this range in \mbh, \lha, and $z$.
\label{bhmfunc}}
\end{figure*}
%\vskip 2mm
%%%%%%%%%%%%%%%%%%%%%%%%%%%%%%%%%%%%%%%%%%%%%%%%%%%%%%%%%%%%%%%%%%%%%

As a matter of practicality, then, we turn the problem around.  Rather
than attempting to quantify our absolute incompleteness as a function
of \mbh, we simply quantify the range of host galaxy luminosities for
which we might hope to detect a BH of a given mass, \lha, and $z$.
Over narrow ranges in all of these parameters, 
neither the line width nor the galaxy continuum strength changes
dramatically and the completeness ought to be constant (provided the BHs
are drawn from the same host galaxy population).  In such bins, with
uniform completeness, we are able to measure true changes in space
density.  Simulations allow us to isolate ranges of \mbh, \lha, and
$z$ with constant sensitivity to galaxy fiber luminosity.  Note,
however, that in any given interval, we necessarily exclude different
members of the population as a function of \mbh; at the lowest masses
we preferentially exclude those systems in faint hosts, while at the
highest \mbh\ we exclude the higher \lledd\ systems.  Therefore, we
implicitly assume that the distribution of \mbh\ is uniform
independent of both disk luminosity and Eddington luminosity.  With
this approach, however, we need not concern ourselves directly with
host galaxy structure\footnote{At a given luminosity, a wide range of
galaxy morphologies are permitted.  Thus it is still possible to find
significant differences in \sigmastar\ (e.g.,~Greene \& Ho 2006b) and
potentially host galaxy structure (J. E. Greene, in preparation)  as a
function of \mbh\ for the SDSS-selected samples of low-mass systems.}
per se, but solely the luminosity.  The total host galaxy luminosity
must be high enough that the source is spectroscopically targeted, while
the fiber luminosity must be low enough to allow detection of the
broad line.

We investigate three mass regimes (\mbh=$10^{5.5}-10^{6.4}$,
$10^{6}-10^{7}$, and $10^{6.5}-10^{7.5}$~\msun), choosing optimal $z$ and
\lha\ ranges for each.  Our procedure is described in most detail for the
lowest (and most challenging) mass bin, and then results are presented for
all three.  In the first bin, we are fundamentally limited by
the total number of objects.  Therefore, we are forced to use the
lowest possible redshift bins: $z = 0.05-0.07$ and $z = 0.07-0.1$.  These are
bins with $\geq 10$ objects per bin for the most part; at still lower
$z$ a prohibitively large range in distance is needed to populate each
bin.  In terms of \lha, the highest luminosity is set by the Eddington
luminosity of the lowest mass bin, in this case $10^{41}$~\lum\ for
a BH with mass $10^{5.5}$~\msun, while the lowest luminosity is set by
the paucity of lower-luminosity objects ($10^{40.5}$~\lum).

Simulations allow us to verify that the selection probability is
indeed independent of host galaxy fiber luminosity.  We make
artificial spectra in the appropriate \mbh\ and \lha\ range, with
fiber galaxy luminosities spanning $-14 \leq M_{B} \leq -22$.  The
galaxy continuum is modeled as a single stellar absorption-line system,
constructed from the eigenspectra of Yip \etal\ (2004), and the S/N is
varied to correspond to typical SDSS spectra over the redshift range
of interest.  Five realizations are made for each galaxy luminosity
and S/N, and each spectrum is run through our full detection
algorithm.  For those with detectable broad \halpha, we then
investigate whether the galaxy luminosity is sufficient for
spectroscopic targeting in the first place.  Recall that this limit
depends on total (rather than fiber) luminosity, but there is not a
one-to-one conversion from fiber to total luminosity; it depends on galaxy
morphology and redshift rather strongly.  Therefore, we place an upper
limit on the total galaxy luminosity by insisting that the fiber
luminosity account for no less than $20\%$ of the total galaxy
luminosity (as motivated by the observed range shown in Fig. 9 of
Tremonti \etal\ 2004).  Over the entire range of galaxy luminosities
we explore, a non-zero detection fraction results only for fiber
luminosities in the range $-16<M_B<-18$, but the detection fractions
at a given host luminosity are very constant across the mass range of
interest, as shown in Figure 8.

In Figure 9{\it a} we show the resulting mass functions for the 
two different redshift bins.  Visually, it appears that the space density 
is truly falling at low mass.  To quantify the significance of this result,
we perform a simultaneous fit to the mass functions in the two 
redshift bins.  We have chosen to use a single power-law 
in this case, 
\begin{equation}
\phi = \frac{\phi^*(M_*)}{M_*} \left(\frac{M}{M_*}\right)^{\alpha}.
\end{equation}
The two redshift bins are constrained to have the same value of
$\alpha$ but are allowed different values of $\phi^*(M_*)$.  An
apparent flat slope in our logarithmic mass function,
$\hat\Phi$(\mbh), would correspond to $\alpha = -1$.  We find a
reduced \chisqr=1.01 and a best-fit slope of $\alpha=-0.6 \pm 0.2$,
corresponding to a $2~\sigma$ detection of a falling space density for
\mbh~$\lesssim 10^{6.5}$~\msun.  If we relax the luminosity criterion
to include all objects with \lha~$=10^{40}-10^{41}$~\lum, the best-fit
slope is $\alpha=-0.7 \pm 0.1$.  We have performed this fit including
the lowest-mass point (\mbh$=10^{5.2}-10^{5.5}$~\msun), but this point
is suspect both because there only 5 objects in the bin, and also
because our maximum \lha\ exceeds the Eddington limit for BHs in this
bin.  However, when this point is removed, the best-fit slope does not
change.  The major outstanding caveats, unavoidable products of our
technique, are that these BHs must both be highly active and live in
fairly luminous host galaxies.

We repeat the above analysis for two other mass intervals.  For the
first, \mbh~$=10^{6}-10^{7}$~\msun, we employ a luminosity interval of
\lha~$=10^{40.5}-10^{41.7}$~\lum\ and, because there are many
more objects in this mass range, three redshift intervals,
$z=0.05-0.08$, $0.08-0.11$, and $0.11-0.14$.  The fitting results are
shown in Figure 9{\it b}, and, with a best-fit $\alpha = -0.72 \pm
0.06$, are consistent with the slope found above.  In this case, we
have a $5~\sigma$ detection of decreasing space density below \mbh$<
10^7$~\msun.  However, it should be noted that the best fit has
\chisqr~$=2.7$, suggesting that this is not a terribly good model.  While
we might achieve an improved fit with a broken power-law model, this
hardly seems justified for such a limited range in BH mass.  Finally,
we turn to the break at high mass, using the mass bin
\mbh~$=10^{6.5}-10^{7.5}$~\msun, and the luminosity bin
\lha~$=10^{41}-10^{42}$~\lum.  Again we are able to utilize three
redshift bins.  Because of the bright magnitude limit of the SDSS, we
are forced to go to higher redshift for completeness in the highest
mass bins here, $z=0.10-0.13$, $0.13-0.16$, and $0.16-0.19$.  In this mass
bin, the mass function begins to turn over: we find a best-fit $\alpha
= -1.13 \pm 0.04$, corresponding to a $3~\sigma$ detection of a
falling space density above \mbh~$\gtrsim 10^{6.5}$~\msun.  Again, the
single power-law model is not a fantastic fit (\chisqr=3.4), so it is
difficult to say when the turnover actually occurs.  If we repeat the
fit without the highest mass bin, the fit apparently improves
(\chisqr~$=1.6$) and the slope steepens, $\alpha = -1.40 \pm 0.05$.  Even
if we neglect the top two or three mass bins, we find the same result.
Therefore, we cannot exactly localize the break; it occurs somewhere in
the range $10^{6.5}-10^{7.5}$~\msun.

\subsection{Comparison with Heckman \etal\ (2004)}

For the first time, we are able to compare narrow- and broad-line AGN
mass functions rather than just luminosity functions (Fig. 10).  The
narrow-line objects come from the SDSS database of Kauffmann \etal\
(2003), for which BH masses are inferred from the stellar velocity
dispersion (Heckman \etal\ 2004; we actually use the most recent
sample from DR4).  Reliable BH masses extend no
lower than $10^{6.3}$~\msun\ due to the resolution limit of the SDSS
($\sigma \approx 71$~\kms; e.g.,~Heckman \etal\ 2004).  Volume
weights depend only on galaxy luminosity, since they were selected
from the main galaxy sample.

%%%%%%%%%%%%%%%%%%%%%%%%%%%%%%%%%%%%%%%%%%%%%%%%%%%%%%%%%%%%%%%%%%%%
%%BoundingBox: 
\begin{figure*}
\vbox{ 
\vskip -0.1truein
\hskip 0.05in
\psfig{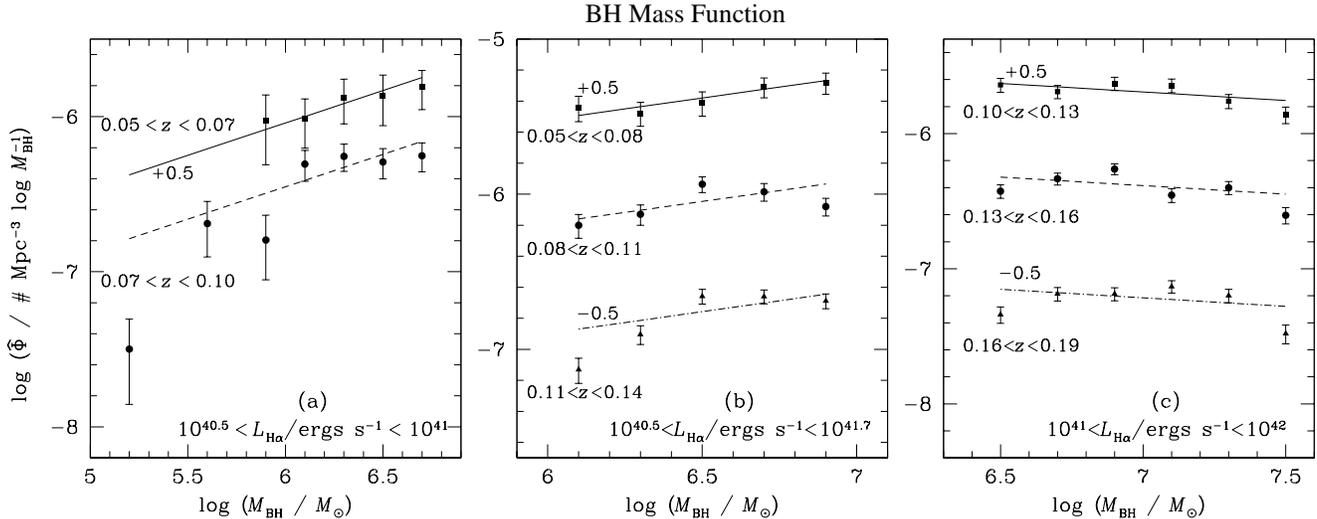}
}
%\vskip -0mm
\figcaption[]{
Fits to the BH mass function in narrow ranges of \mbh, \lha, and $z$,
as indicated.  These windows were selected to have uniform sensitivity
to galaxy fiber luminosity, so that the measured slopes represent
physical changes in space density.  We perform simultaneous power-law
fits (allowing different amplitudes but fixing the slopes between
bins) to the redshift bins shown in each panel.  We clearly detect a
break in the mass function in the mass range $10^{6.5}-10^7$~\msun,
although the exact location is not clear from these data.  Note also
that we are only sensitive to active galaxies with a narrow range of
fiber luminosities, which may bias these results (see text). 
\label{bhmfunc}}
\end{figure*}
%%%%%%%%%%%%%%%%%%%%%%%%%%%%%%%%%%%%%%%%%%%%%%%%%%%%%%%%%%%%%%%%%%%%%

It is clear that the space density of narrow-line AGNs in the Heckman
\etal\ sample is higher than that of the broad-line sample by an order
of magnitude.  However, it is crucial to remember that the selection
effects for the two samples are radically different.  In particular,
narrow emission lines may be detected to significantly lower
luminosities than broad lines.  Therefore, while the typical Eddington
ratio in our sample is \lledd\ $\approx 0.1$, the typical \lledd\ for
a $10^7$~\msun\ BH in the Heckman \etal\ sample is closer to
\lledd$\approx 0.04$ (see their Fig. 3).  It would be much more
informative to compare the broad- and narrow-line objects in matching
mass and luminosity bins.  However, to do so properly requires a
consistent measure of AGN luminosity across the two samples.
Typically, the \oiii\ line is used (e.g.,~Zakamska \etal\ 2003), but
for broad-line objects there may be significant contamination to this
line from broad \feii\ emission (e.g.,~Greene \& Ho 2005b).  Since we
have not properly subtracted the \feii, a more detailed comparison
must await future work.

\section{Discussion}

\subsection{Active Fraction or Duty Cycle}

We have constructed a BH mass function for the local Universe, using
broad-line AGNs and virial BH mass estimates.  By looking at very
narrow ranges in redshift and AGN luminosity, we have detected a break
at $M_{\rm BH} \approx 10^{6.5}-10^7$~\msun.  We now investigate what
fraction of BHs are active as a function of BH (spheroid) mass, by
computing the inferred shape of the inactive BH mass function, using
the formalism of Marconi \etal\ (2004; see also Yu \& Tremaine 2002; 
Shankar \etal\ 2004; McLure \& Dunlop 2004).
Statistically speaking, the fraction of currently active systems may
be interpreted as a duty cycle for BHs in that mass range.

Marconi \etal\ compute consistent inactive mass function from a
variety of optical and near-infrared surveys and so for simplicity we
use the Kochanek \etal\ (2001) $K$-band galaxy luminosity function.
The galaxy luminosity and BH mass functions are linked through the
$K$-band \mlb\ relation of Marconi \& Hunt (2003),
assuming an intrinsic scatter in that relation of 0.3 dex and using 
Monte Carlo realizations to incorporate measurement uncertainties.
The resulting uncertainties (see their Fig. 2{\it b}) are
relatively constant for \mbh$\lesssim 10^8$~\msun, and have an
amplitude of $\sigma \approx 0.2$ dex, although there are a variety of
systematic uncertainties that preferentially impact the low-mass
regime (for comparison, the Shankar et al. mass function is $\sim 0.5$ dex
higher than Marconi \etal\ at \mbh$=10^7$\msun).  At high mass, 
the mass function is well-constrained; the luminosity of elliptical galaxies
may be translated directly into BH mass.  On the other hand, to
properly translate between galaxy luminosity and BH mass at low
luminosities requires knowledge of the (poorly constrained) luminosity
function of bulges.  For the moment, we simply adopt the fiducial
values from Marconi \etal, who assume morphological fractions ($m_{\rm
bulge} - m_{\rm total}$) from Fukugita \etal\ (1998) of 11\%
elliptical, 21\% S0, 43\% Sab, and 19\% Scd and bulge-to-disk values
from Aller \& Richstone (2002; derived from the data of Simien \&
de~Vaucouleurs 1986) of $0.64\pm 0.30$ for S0, $1.46\pm0.56$ for Sab,
and $2.86\pm0.59$ for Scd galaxies.  We show the resultant
distribution of inactive BH mass as a dashed line in Figure 11{\it a}.
For reference, the contribution from early- and late-type spirals
within the adopted formalism are shown in long and short dot-dash
lines, respectively.

The active fraction clearly depends strongly on BH mass.  The
differences in shape between active and inactive systems is
highlighted in Figure 11{\it a}, where the active mass function is
boosted by a factor of $\sim 200$ to overlay the inactive mass function
derived as described above.  (Because the low-mass slope is
particularly uncertain, we also show the best-fit slope to the
lowest-mass and redshift bin from \S 4.4 [{\it dotted}]). In the
bottom panel we show the ratio of each subpopulation with the inactive
mass function.  BHs with masses $\sim 10^7$~\msun, living in $\sim
10^{10}~M_{\odot}$ spheroids are active $\sim 0.4$\% of the time.  At
higher BH mass the number of active BHs drops quickly; local massive
BHs are not radiating at substantial fractions of their Eddington
limit.  The same behavior was seen by Heckman \etal\ (2004) for
narrow-line AGNs.  It is reassuring that we find consistency with the
results of Heckman \etal, despite our differing (indirect) means of
measuring \mbh.  The Heckman \etal\ result, and now ours, may be seen
as confirmation of the suggestion, based on AGN luminosity functions,
that lower-mass objects build their mass at low redshift, while the
most massive BHs completed their growth at high redshift (``cosmic
downsizing''; Cowie \etal\ 1996; Ueda \etal\ 2003; Steffen \etal\
2003).

There is an important caveat that deserves discussion.  While Marconi
\etal\ (as well as Shankar \etal\ and McLure \& Dunlop) find good
agreement between inactive mass functions derived from galaxy
luminosity and \sigmastar\ functions, Tundo \etal\ (2007) report that
the mass function based on \sigmastar\ distributions is systematically
lower than that based on galaxy luminosity functions.
%%%%%%%%%%%%%%%%%%%%%%%%%%%%%%%%%%%%%%%%%%%%%%%%%%%%%%%%%%%%%%%%%%%%
%%BoundingBox: 
%\begin{figure}
%\vbox{ 
%\vskip -0.1truein
\hskip -4mm
\psfig{file=bhmfunct1t2_all.epsi,width=0.45\textwidth,keepaspectratio=true,angle=0}
%}
\vskip -0mm
\figcaption[]{
Comparison between the BH mass function of broad-line 
({\it filled symbols}; this paper, truncated at $10^6$~\msun\ for 
consistency with the narrow-line AGN sample) and narrow-line 
({\it dashed line}; Heckman \etal\ 2004) AGNs.  Mass function for broad-line 
objects as described above.  Volume corrections for the narrow-line objects 
are based solely on the galaxy color and luminosity.  The space density 
of narrow-line objects in their sample is significantly higher than 
what is seen for broad-line objects, due to the relative ease of detecting
high-EW narrow lines.
}
%\label{fwhm}}
%\end{figure}
\vskip 5mm
%%%%%%%%%%%%%%%%%%%%%%%%%%%%%%%%%%%%%%%%%%%%%%%%%%%%%%%%%%%%%%%%%%%%%
\noindent
According to Bernardi \etal\ (2007), the \msigma\ calibration sample
(e.g.,~Tremaine \etal\ 2002) has a biased Faber-Jackson (1976)
relation compared to the SDSS, in the sense that the galaxies have
lower luminosities at a given \sigmastar\ (or higher \sigmastar\ at a
given $L$).  One consequently infers a lower BH mass function from the
\msigma\ relation than from the \mlb\ relation.  Bernardi \etal\
(2007) present a model in which the \msigma\ relation provides the
more reliable estimator of BH mass density.  If (and this is by no
means certain) they have properly modeled the bias, then the true
inactive BH mass density should be $\sim 0.5$ dex lower at
\mbh$=10^9$~\msun, and the active fraction declines less steeply with
mass than we report here.  At the present time, given the systematic
differences between authors and estimators, we conservatively assign a
factor of 3 uncertainty to our active fraction across all mass
bins.

It is also worth noting that the active fraction depends sensitively
on the depth of the survey.  The Palomar spectroscopic survey of
galaxies, which is significantly more sensitive to low-luminosity
\halpha\ lines, finds an active fraction of $\sim 60$~\% in nearby,
bulge-dominated galaxies (Ho \etal\ 1997b).  Despite the limited
dynamic range in our sample, the number density of sources at a given
BH mass increases at lower Eddington ratio, modulo incompleteness
(Fig. 11{\it b}).

Aside from the known decrease of active fraction at high BH mass, we
additionally see a decline in active fraction for BHs $\lesssim
10^{6.5}$~\msun, but unlike the case of massive BHs, we cannot clearly
interpret the decline.  Taken at face value, the observation suggests that
as we move to masses below the peak, the duty cycle declines.
Unfortunately, both the active and inactive mass functions are highly
uncertain for masses $<10^{6.5}$~\msun.  Ultimately, we are limited by
our knowledge of (a) the shape of the luminosity function, (b) the
conversion between galaxy luminosity and BH mass for systems that
contain BHs, and (c) the occupation fraction of BHs in dwarf galaxies.
We discuss the potential magnitude and impact of each of these
effects below.

First of all, the luminosity function of Kochanek \etal\ (2001) is
measured only to a luminosity of $M_K = -20$ mag, or a BH mass of
$10^{6.4}$~\msun\ for an elliptical galaxy, or $10^{5.1}$~\msun\ for
an Scd galaxy.  Thus, the entire low-mass regime involves some
extrapolation.  Deeper luminosity functions do exist (e.g.,~Blanton
\etal\ 2005), but we are still left with the highly uncertain task of
converting between an observed luminosity function and a BH mass
function. The morphological fractions adopted by Marconi \etal\ (2004;
from Fukugita \etal\ 1998) are relatively well-determined for E/S0
galaxies.  They are adopted from the morphology-density study of
Postman \& Geller (1984), based on the CfA redshift survey (Huchra
\etal\ 1983; morphologies from de Vaucouleurs \etal\ 1976 and Nilson
1973), which separated galaxies into E, S0, and spiral using visual
classification.  On the other hand, the fractions of Sab and Scd
galaxies are derived in a less-direct fashion.  Tinsley (1980) took
the observed color and apparent magnitude distributions from the
photographic plate study of Kirshner \etal\ (1978) and solved
simultaneously for the morphological fractions and redshifts that
reproduced the observations, using broad-band color distributions from
Pence (1976).  Because of the degeneracies inherent in photometric
redshifts based on two broad-band colors\footnote{For instance,
Fukugita \etal\ (1995) find that the $K$-corrections for Sa--Sc
galaxies at $z\lesssim 0.3$ differ by $\delta z \lesssim 0.1$, in $B$
or $V$.  Degeneracies between metallicity and age only add to the
uncertainties.}, and the limited sample size ($\sim$ 800 galaxies),
this leads to significant uncertainties in the bulge fractions for
spiral galaxies.

Ideally, we would like a direct measurement of bulge (rather than
total galaxy) luminosity.  D\v{z}anovi\'{c} \etal\ (2007; see also
Tasca \& White 2007) present spheroid luminosity and mass functions based
on two-dimensional image decomposition of $\sim 9000$ SDSS galaxies.
The resulting BH mass function is consistent with a decrease in BH
mass density of $\sim 0.5$ dex between \mbh~$=10^7$~\msun\ and
\mbh~$=10^5$~\msun.  However, the conversion from spheroid mass to BH
mass relies on extrapolation of BH-bulge relations that have been
calibrated only for higher masses.  The host galaxies of low-mass BHs
(\mbh~$< 10^6$~\msun) may not even contain bulges.  Indeed,
neither of the two best-studied low-mass BHs (\mbh~$\approx
10^5$~\msun)---NGC 4395 (Filippenko \& Ho 2003) or POX 52 (Barth
\etal\ 2004)---contains a classical bulge.  The former is a late-type,
bulgeless spiral, which lies on the low-mass extrapolation of the
\mlb\ relation only if one considers the luminosity of its nuclear
star cluster.  POX 52 is a dwarf spheroidal galaxy\footnote{Such
objects are commonly referred to as dwarf elliptical galaxies in the
literature.  However, because their structure is quite different from
that of classical elliptical galaxies, we prefer to refer to them as
dwarf spheroidal systems.}, and from the total luminosity of the host
one would predict a BH mass 50 times higher than its virial mass or
\sigmastar\ suggests.  A similar trend is seen for the SDSS sample of
Greene \& Ho (2004; J.~E.~Greene \etal, in preparation).  It is
possible that these more distant systems also obey an \mlb\ relation
with a nuclear star cluster, and POX 52 may show evidence of
nucleation from \hst\ observations (C.~J. Thornton \etal, in
preparation), but in general it is not possible to predict nuclear
cluster luminosity from galaxy luminosity (although the two may be
correlated; Ferrarese \etal\ 2006; Rossa \etal\ 2006; Wehner \& Harris
2006).  In short, a conversion from galaxy luminosity to BH mass is
presently unconstrained for BHs with masses \mbh$< 10^6$~\msun.

%%%%%%%%%%%%%%%%%%%%%%%%%%%%%%%%%%%%%%%%%%%%%%%%%%%%%%%%%%%%%%%%%%%%
%%BoundingBox: 
\begin{figure*}
\vbox{ 
\vskip -0.1truein
\hskip 0.2in
\psfig{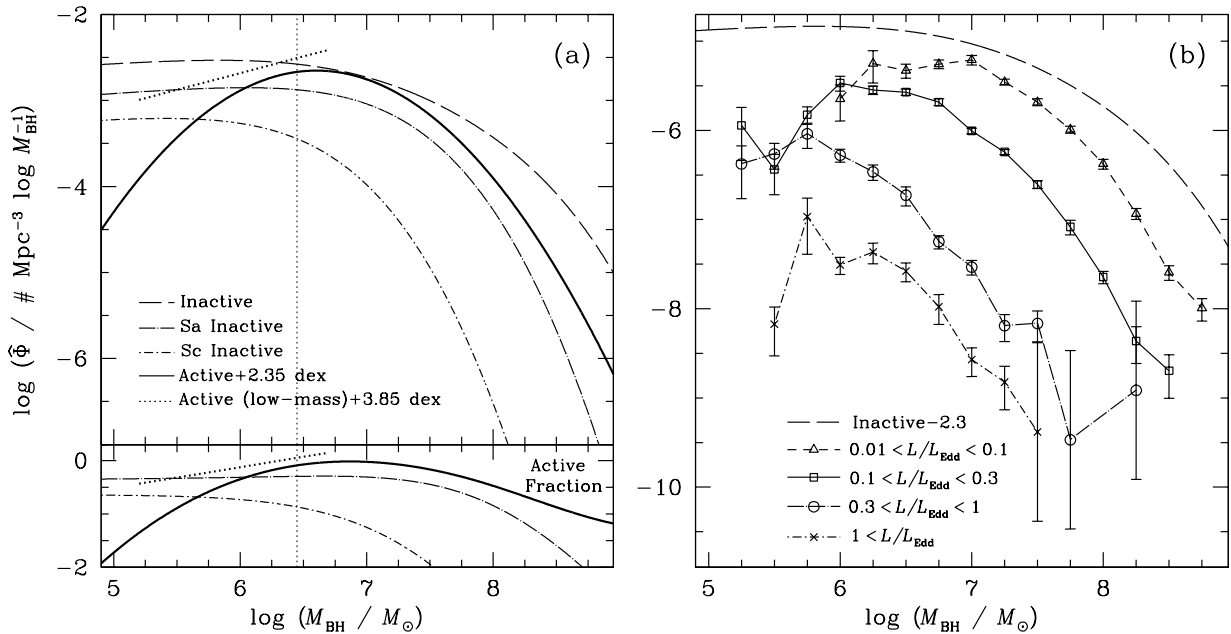}
}
\vskip -0mm
\figcaption[]{
({\bf \it a})
Comparison of the inactive BH mass function with the broad-line BH
mass function.  The inactive mass function is derived from the galaxy
luminosity function of Kochanek \etal\ (2001), using the formalism of
Marconi \etal\ (2004), including an intrinsic dispersion of 0.3 dex in
the relation between galaxy luminosity and BH mass.  We show the
log-normal fit to the active BH mass function offset by 2.35 dex ({\it
solid}) and the best-fit power law to the lowest-mass, lowest-$z$ bin
(Fig. 9{\it a}) to demonstrate the uncertainty in slope and break
position at the low-mass end.  The ratio between the total inactive
mass function and each population is shown in the bottom panel.  The
vertical line indicates the point at which we begin to use an
extrapolation of the luminosity function.
({\bf \it b})
Comparison of the inactive BH mass function ({\it dashed}) with the active 
mass function in different Eddington ratio bins ({\it solid symbols}).
Bins with $>3$ points are shown.  It is clear that all high-mass BHs 
are radiating at very low Eddington ratios.
\label{act_inact}}
\end{figure*}
%\vskip 5mm
%%%%%%%%%%%%%%%%%%%%%%%%%%%%%%%%%%%%%%%%%%%%%%%%%%%%%%%%%%%%%%%%%%%%%

Finally, while we believe that most (if not all) bulges host BHs, the
occupation fraction of BHs in late-type galaxies in unknown.
Dynamical measurements in the Local Group for the late-type galaxy M33
(Gebhardt \etal\ 2001) and the dwarf spheroidal galaxy NGC 205
(Valluri \etal\ 2005) show that neither hosts a BH with a mass in
accordance with expectations from the \msigma\ relation.  On the other
hand, the massive globular cluster G1 in Andromeda hosts a BH in
accord with the \msigma\ relation (Gebhardt \etal\ 2002, 2005), and
NGC 4395, POX 52, and the Greene \& Ho (2004) sample, all late-type
galaxies hosting BHs with masses $< 10^6$~\msun, are consistent with
the low-mass extrapolation of the \msigma\ relation (Filippenko \& Ho
2003; Barth \etal\ 2004, 2005).

In light of these many uncertainties, it is difficult to interpret the
apparent decrease in active fraction at low mass.  The most
straightforward interpretation may be that the shape of the active
mass function actually traces the shape of the bulge luminosity
function at low luminosity.  Otherwise, is it is difficult to
understand why late-type spiral galaxies, with a ready gas supply,
nevertheless are preferentially inactive.  The distribution of
activity in the Palomar spectroscopic survey of nearby galaxies
support this hypothesis; Ho \etal\ (1997b) find that Seyfert galaxies
(i.e., objects with relatively high accretion rates) are predominantly
found in early-type spiral galaxies (Sa--Sb).  In the future, with the
availability of more complete measurements of spheroid luminosity
functions, we may use our observations of active fractions to measure
the mass at which the BH occupation fraction departs from unity.

\subsection{Connection to Higher Redshift}

Broad-line AGN activity at the present day is dominated by $\sim
10^7$~\msun\ BHs radiating at $\sim 10\%$ of their Eddington limits
(Fig. 11{\it b}).  During the quasar epoch, $1<z<3$, BH mass growth is
dominated by optically bright, near-Eddington growth of massive BHs
(e.g.,~Vestergaard 2002) and this growth basically accounts for most
of the total BH mass density observed today (e.g.,~Yu \& Tremaine
2002).  Even as one moves to lower luminosities, BH growth appears to
be dominated by near-Eddington accretion, at least for massive BHs at
high redshift (Kollmeier \etal\ 2006).  In general, this picture sits
comfortably with the observation that massive elliptical galaxies
(hosts of the most massive BHs) formed their stars rapidly at high
redshifts (e.g.,~Bower \etal\ 1992; Trager \etal\ 2000; Thomas \etal\
2005).  It is the redshift range between the present and $z\approx 1$
that remains poorly constrained.  From the point of view of galaxy
evolution, there appears to be a significant (factor of 2) increase in
the mass density of spheroids over this interval (e.g.,~Brown \etal\
2007; Faber \etal\ 2007).  It would be interesting to know whether any
commensurate BH growth occurred.

There are tantalizing hints.  Optical QSO luminosity functions show
that massive BHs are not radiating at high Eddington ratio below $z
\approx 1$ (e.g.,~Richards \etal\ 2006), but do not constrain the
evolution of low-luminosity sources.  X-ray luminosity functions,
built from very deep \chandra\ and \xmm\ pointings, are able to probe
significantly fainter luminosities (see Brandt \& Hasinger 2005 for a
recent review).  It is clear that the space density of low-luminosity
($L_{\rm X} \approx 10^{42}-10^{43}$~\lum), X-ray--selected sources
peaks at considerably lower redshift ($z \approx 0.7$) than
higher-luminosity sources.  Unfortunately, multiple scenarios for BH
growth result in similar luminosity evolution.  At one extreme, one
may imagine all near-Eddington growth at all masses occurs at high
redshift and then slowly shuts off, while at the other extreme, one
may imagine that the low-luminosity radiation is dominated by the
growth of low-mass BHs, which acquire a significant fraction of their
mass at low redshift.  In reality, some combination of these two
scenarios presumably occurs.  The latter hypothesis, dubbed ``cosmic
downsizing,'' is often preferred in the literature, partially because
galaxies display similar behavior (e.g.,~Cowie \etal\ 1996; Barger
\etal\ 2005).

The first concrete evidence in support of downsizing came from the
Heckman \etal\ (2004) study, which showed unambiguously that local
accreting BHs are typically an order of magnitude less massive than
the typical local inactive BH.  The present study finds the same
result.  However, we still must ask whether X-ray--selected AGNs at
$z=0.7$ are really the same population as local optically selected
sources.  The story may not be so straightforward.  Netzer \&
Trakhtenbrot (2007) find that the typical BH mass and Eddington ratio
of optically selected AGNs both increase significantly from the
present time to $z=0.7$, so that what we measure locally may no longer
apply at that epoch.  The other intriguing, but not definitive,
evidence comes from recent optical studies of the host galaxies of
X-ray--selected AGNs.  It seems that the majority of sources are found
in massive, relatively red galaxies (e.g.,~Colbert \etal\ 2005; Nandra
\etal\ 2007; see also Barger \etal\ 2005).  The clustering properties
of X-ray--selected AGNs also suggest that the hosts are typically more
massive than optically-selected AGNs at the same redshift (e.g.,~Coil
\etal\ 2007, Miyaji \etal\ 2007, Coil, A.~L. private communication).
Assuming that the \msigma\ relation is largely in place at $z \approx
0.7$ (Peng \etal\ 2006; Salviander \etal\ 2007; but see also Treu
\etal\ 2004; Woo \etal\ 2006), the observed galaxy luminosities
correspond to BHs with masses $> 10^8$~\msun, casting doubt on the
cosmic downsizing scenario for this population.

Taken at face value, the fact that X-ray--selected samples are found
in very luminous host galaxies suggests that low-luminosity, hard
X-ray--selected sources consist predominantly of massive BHs in a
low-Eddington state.  These sources are known to be an intrinsically
hard population with no big blue bump (Ho 1999), and if they have
broad lines, they will be extremely low contrast, and thus difficult
to detect.  The conjecture that low-Eddington sources dominate the
hard X-ray population thus accounts for both the general lack of
apparent broad emission and the generic hard spectral shape of the
X-ray background (see additional arguments in Shen \etal\ 2007).  On
the other hand, low-luminosity, soft X-ray--selected sources, most
likely closely linked to optically selected broad-line AGNs, also have
a peak space density at low redshift (e.g.,~Hasinger \etal\ 2005).  At
the present time, it is unclear whether the massive host galaxies
measured in the studies above are representative of the population as
a whole.  An unbiased comparison of the Eddington ratio distributions
of the soft and hard X-ray--selected samples is needed.

A rather different approach to constraining the evolution of BH mass
density is taken by Merloni (2004).  He constructs a BH mass function
using constraints from the joint X-ray and radio luminosity function,
in combination with an empirical relation between BH mass and X-ray
and radio luminosity (the ``fundamental plane''; e.g.,~Merloni \etal\
2003).  He finds fast evolution; by $z=0.6$ the
characteristic accreting BH mass has increased from $10^7$~\msun\ to
$10^8$~\msun.  However, his analysis also tends to favor low-Eddington
sources, and indeed his characteristic $L_{\rm X}/L_{\rm Edd}$ never
exceeds $1\%$.  Ideally, work like that presented here may be extended
to higher redshift to directly address the distribution of BH mass at
intermediate redshift.

\section{Summary}

We present the first measurement of the local BH mass function for
broad-line active galaxies.  Our sample of $\sim 9000$ galaxies is
drawn from the Fourth Data Release of the Sloan Digital Sky Survey,
based on the presence of broad \halpha\ emission.  Our \halpha\
luminosity function is found to be consistent with previous
measurements, and with the local soft X-ray luminosity function.
Using standard scaling relations between AGN luminosity and line
width, we derive BH masses for the entire sample.  Much like the
Heckman \etal\ (2004) sample of narrow-line AGNs, the typical BH in
our sample has a mass of \mbh\ $\approx 10^7$~\msun\ and an Eddington
ratio of \lledd\ $\approx 10\%$.  Although we are highly incomplete for
low-Eddington ratio systems, by looking at very narrow bins in \mbh,
\lha, and $z$ we find clear evidence for a true turnover in space
density below \mbh\ $\approx 10^{6.5}-10^7$~\msun.  Compared to the
inferred shape of the mass function of inactive BHs, the mass function
of active BHs falls significantly both above and below this
characteristic break mass.  The dearth of active massive BHs is a
familiar result---massive BHs are mostly quiescent in the local
Universe.  The decreasing space density at low BH mass presumably
reflects the fact that bulge fraction and BH occupation fraction both
decrease in dwarf galaxies.

One of the major motivations of this study was to constrain the shape
of the BH mass function for BHs with masses $<10^6$~\msun, since
broad-line AGN surveys currently have the unique capability to explore
this mass regime.  We have found evidence for a decreasing space
density of active BHs with masses $<10^{6.5}$~\msun, at least for
objects with host galaxies more luminous than $M_{B} \approx -16$ mag.
However, substantial work remains to determine the true demographics
of BHs in low-mass galaxies.  Measurements that isolate the luminosity
and mass functions of spheroids (e.g.,~D\v{z}anovi\'{c} \etal\ 2007)
on the one hand, combined with empirical conversions between spheroid
luminosity and BH mass in the low-mass regime (J.~E.~Greene \etal, in
preparation) on the other, will provide predictions for the shape of
the inactive BH mass function for \mbh$<10^6$~\msun.  At the same
time, alternate search techniques, which are less sensitive to host
galaxy luminosity, are required to discover low-mass, low \lledd\ BHs.
Apart from deeper nuclear optical spectroscopic surveys, deep X-ray
and radio surveys may provide sensitivity to lower-Eddington ratio
systems and eliminate the host luminosity bias suffered by the present
study.  Finally, future time-domain surveys may place limits on the
frequency of flares from the tidal disruption of stars captured by BHs
in dwarf galaxies (e.g.,~Rees 1988; Donley \etal\ 2002; Gezari \etal\
2006; Milosavljevi\'c et al. 2006).

\acknowledgements
We gratefully acknowledge useful conversations with M.~Bernardi,
L.~Hao, J.~P.~Ostriker, J.~A.~Kollmeier, and, particularly,
M.~A.~Strauss.  We appreciate a timely report from the anonymous
referee.  Finally, we thank the entire SDSS team for providing
the fantastic data products that made this work possible.  Support for
J.~E.~G. was provided by NASA through Hubble Fellowship grant HF-01196
awarded by the Space Telescope Science Institute, which is operated by
the Association of Universities for Research in Astronomy, Inc., for
NASA, under contract NAS 5-26555.  L.~C.~H. acknowledges support by
the Carnegie Institution of Washington and by NASA grant SAO 06700600.
Funding for the SDSS has been provided by the Alfred P. Sloan
Foundation, the Participating Institutions, the National Science
Foundation, the U.S. Department of Energy, the National Aeronautics
and Space Administration, the Japanese Monbukagakusho, the Max Planck
Society, and the Higher Education Funding Council for England. The
SDSS web site is {\tt http://www.sdss.org/}.

\appendix

\section{A. Measurement Uncertainties}

We describe a suite of simulated galaxy spectra that we have built to
investigate the reliability of our BH mass measurements over the
\halpha\ luminosity range of interest, for SDSS-quality data.  We
investigate our ability to recover the input \fwha, \lha, and \mbh\ as
the \halpha\ flux, galaxy continuum, and S/N of the spectra change.
We also justify our particular choice of detection thresholds in
rms-normalized flux and EW(\halpha).  At the same time, we
investigate the importance of factors such as strength of the narrow
lines, shape of the stellar continuum, and the slope of the AGN
continuum.  The basic grid of simulations spans a range of
$10^4<$\mbh/\msun$<10^9$, $0.01<$\lledd$<3$, and $5<$S/N$<50$.  As we
have noted previously, the level of galaxy continuum may play a
crucial role in our ability to measure the broad \halpha\ component at
all.  We have decided to assign the galaxy luminosity based roughly on
scaling relations between the BH and the surrounding galaxy.  While
there is large scatter in such scaling relations, we attempt to span a
reasonable range of galaxy luminosities, so that at least we are
simulating realistic conditions.

All of the spectral properties are derived from the chosen \mbh,
\lledd, and S/N.  In terms of the AGN features, the continuum
luminosity and \fwha\ are determined using the formulae
presented in Greene \& Ho (2005b).  A small fraction ($10\%$) of the
total broad emission luminosity is placed in a very broad component,
since such broad wings are often seen in actual spectra.  The AGN
continuum shape is given slopes of $\beta=1.5, 1.0$, and 0.5, where
$f_{\beta}\propto \lambda^{-\beta}$.  Using the Tremaine \etal\ (2002)
\msigma\ relation, we calculate the width of the  narrow-line region
components, assuming that the velocity dispersion of the narrow
lines is equal to that of the bulge (e.g.,~Nelson \& Whittle 1996;
Greene \& Ho 2005a), while the  narrow-line flux is fixed using the empirical
relation of Zakamska \etal\ (2003), which relates \loiii\ to \lf.
Finally, all lines are convolved to the typical velocity resolution of
the SDSS, $\sigma \approx 71$ \kms\ (e.g.,~Heckman \etal\ 2004).

In terms of the host galaxy, we again assume the \msigma\ relation
holds, and associate a stellar velocity dispersion and bulge
luminosity with each BH.  For elliptical galaxies, this is the entire
story, but at low \mbh, the hosts are typically spiral galaxies.
Therefore, we need a way to estimate the total galaxy luminosity based
on the bulge properties.  The \msigma\ relation of Tremaine \etal\
(2002) links each input BH with a \sigmastar.  Then, using the
relation between \sigmastar\ and maximum circular velocity \vc\
(e.g.,~Ferrarese 2002; Baes \etal\ 2003; Pizzella \etal\ 2005) to
obtain \vc, the Tully-Fisher relation (Tully \& Fisher 1977) may be
used to estimate a total galaxy luminosity for a spiral host.  In
particular, we use the best-fit Tully-Fisher relation from Masters
\etal\ (2006) for their ``in$+$'' sample, and we neglect variations in
the relation with Hubble type.  We incorporate the scatter in both the
\vc-\sigmastar\ relation and the Tully-Fisher relation by perturbing
each fit parameter by a log-normally distributed deviate drawn from
the best-fit distribution.  For a given BH, we must decide whether to
place it in a spiral or elliptical host galaxy.  There is, of course,
a limit to how bright a spiral galaxy may be and, conversely, to the
faintest elliptical galaxies.  For the latter, we adopt the luminosity
of M32 ($M_B = -15.8$ mag; Tremaine \etal\ 2002), while for the former
we adopt a limit of $M_I > -24$ mag, which represents the brightest
galaxies in the Masters \etal\ sample.  We note that M32 is fainter
than POX 52, which is a dwarf spheroidal galaxy known to host a BH of
mass $\sim 10^5$~\msun\ (Barth \etal\ 2004).  This is because dwarf
spheroidal galaxies are more luminous than elliptical galaxies at the
same \sigmastar\ (e.g.,~Geha \etal\ 2003).  If the calculated bulge
luminosity is too faint, the object is assigned only a spiral host,
and conversely if the luminosity is bright enough, the object is
assigned an elliptical host.  BHs in the mass range of
$10^{6.5}-10^{8}$~\msun\ are typically assigned both, which is in good
agreement with observations.  The stellar continuum shape is provided
by the eigenspectra presented in Yip \etal\ (2004).  For the majority
of the simulations, we use a pure absorption-line spectrum typical of
an elliptical galaxy, by combining their first three eigenspectra with
weights [1,1,0.9].  We have investigated the impact of younger stellar
populations, using weights [0.1,0.0,$-$0.5] to represent a typical Sc
galaxy, and [0.0,0.0,$-$0.5] to represent a post-starburst system.  We

%%%%%%%%%%%%%%%%%%%%%%%%%%%%%%%%%%%%%%%%%%%%%%%%%%%%%%%%%%%%%%%%%%%%
%%BoundingBox: 
%\begin{figure*}
%\vbox{ 
%\vskip -0.1truein
\hskip 0.5in
\psfig{file=masslumerr_nsigeqw.epsi,width=0.7\textwidth,keepaspectratio=true,angle=0}
\vskip -0mm
\figcaption[]{
Deviation $\Delta=$~(out$-$in)/in, from the measured properties \lha,
\fwha, and the inferred \mbh\ for all simulations.  We show the
deviations in ({\it left}) logarithmic bins of \halpha\ flux normalized by the
local rms and ({\it right}) EW(\halpha). The entire
sample in each bin is shown as an open circle, while the sub-sample
with the joint flux and EW cuts are shown as open squares; they only
differ at and below the detection thresholds.  Objects that have been
rejected by our initial detection are not included, but those below
our detection threshold are, so that we may justify our choices.  The
detection thresholds are indicated by dotted lines.
\label{fwhm}}
%\end{figure*}
\vskip 5mm
%%%%%%%%%%%%%%%%%%%%%%%%%%%%%%%%%%%%%%%%%%%%%%%%%%%%%%%%%%%%%%%%%%%%%
\noindent
find that changes in stellar populations do not significantly impact
our derived masses; the PCA method of Hao \etal\ (2005a) is quite
robust for our purposes.

Each artificial spectrum is run through our entire detection
algorithm.  Continuum subtraction is performed using PCA, the \halpha\
detection is performed, and then the line fitting is performed to see
if the broad lines are detectable at this S/N.
Objects that would be rejected by our initial \halpha\ detection
algorithm are not included in the investigation of parameters shown
below, but we do not remove objects below our detection threshold,
since it is our goal to demonstrate that we have chosen reasonable values.  We 
quantify the dependence of the measurements on normalized flux and EW(\halpha) 
by plotting as a figure of merit the fractional deviation $\Delta X \equiv
(X_{\rm out}-X_{\rm in})/X_{\rm in}$.  Specifically, we calculate $\Delta$\lha,
$\Delta$\fwha, and $\Delta$\mbh.  Note that these deviations are
plotted in linear (not logarithmic) space.  A factor of 2 error in
\mbh\ corresponds to $\Delta$\mbh=1.  We show in Figure A12 the
dependence of $\Delta X$ on the rms-normalized flux and EW(\halpha)
for each parameter.  The detection threshold is noted as a dotted line
in each panel.  It is quite clear from these figures that below our
chosen detection thresholds we incur completely unacceptable
uncertainties in the derived parameters.  We have run smaller sets of
simulations to test the importance of both the narrow-line region
strength and the underlying galaxy continuum shape, and we find that
the deviations in \lha, \fwha, and \mbh\ are constant as a function of
these parameters.  If anything, low-mass BHs are slightly more likely
to be included in our sample when the narrow-line strength is high
compared to our fiducial choice.

\section{B. Maximum Likelihood}

While the \vvmax\ method used in \S4.1 is extremely intuitive, 
the resulting luminosity functions may be biased due to 
local inhomogeneities in the distribution of galaxies, as well as 
peculiar motions within the Local Supercluster
(e.g.,~Efstathiou \etal\ 1988).  In the maximum likelihood method, 
rather than parameterizing the AGN selection function as a maximum 
observable volume, each source is assigned a probability 
of detection as a function of \halpha\ luminosity, $p_i$(\lha).  Then, 
the likelihood that we observe our particular sample may 
be expressed as 
\begin{equation}
\mathcal{L} = \frac{\prod_{i}p_i(L_{i}) \Phi(L_i) dL_i}
{\int p_i(L_i) \Phi(L) dL}. 
\end{equation}
Our goal is to maximize this function, or equivalently minimize 
$S \equiv -2~{\rm ln}~\mathcal{L}$.  Rather than assume a particular form 
for the underlying luminosity function, we derive a stepwise-constant, 
non-parametric function.  In detail, we follow the iterative procedure 
outlined in Blanton (2000; see also Koranyi \& Strauss 1997; 
Hao \etal\ 2005b).

The primary challenge is in deriving the selection function, or the
probability of observing a given object as a function of luminosity.
We attempt to follow closely the technique of Hao \etal\ (2005b) for
ease of direct comparison.  Given the flux limits of the SDSS, each
source has both a minimum and a maximum luminosity at which we may
observe it.  As \lha\ increases, we increase the power-law luminosity
of the AGN according to the relation of Greene \& Ho (2005b), and the
maximum luminosity is reached when the source reaches either its
Eddington luminosity, or the bright limit of the SDSS.  The minimum
luminosity is more complicated to calculate.  It is reached either
when the source reaches a magnitude limit, or when the \halpha\ line
is no longer detectable using our algorithms.  For objects targeted as
galaxies, this is relatively straightforward to determine.  At each
luminosity, we calculate the new total magnitudes and corresponding
typical S/N.  The minimum luminosity is reached when the source
reaches our detection threshold (\fha/rms $<$ 200) or the galaxy
magnitude limit.  For the objects targeted as QSOs, the situation is
somewhat complicated by the possibility that the object changes target
selection from QSO to galaxy as the source fades.  The AGN
contribution is estimated from the \halpha\ luminosity.  We adopt a
very ad hoc prescription, in which the object is converted from a QSO
to a galaxy when the AGN accounts for less than $10\%$ of the total
Petrosian magnitude.  In practice, the number we choose does not
affect our results, since in the vast majority of cases the limit is
set by the detection threshold.

Within the limiting magnitudes, the AGN will have different detection
thresholds depending on the contrast of the \halpha\ luminosity and
the S/N.  As the \halpha\ luminosity decreases for a given
source, several important properties of the spectrum will change.  The
continuum level will decrease, of course, but since the BH mass is not
changing, the line shape also must change.  Progressively, as objects
move to lower Eddington ratio, they grow more difficult to find, as
their \halpha\ flux is distributed more and more broadly, and the
overall S/N of the spectrum decreases in parallel.  In terms of
detection efficiency, as we show in Appendix A, the galaxy type and
luminosity only provide secondary corrections to our detection
efficiency, which depends most strongly on line luminosity, line
strength, and S/N.  For that reason, we use our suite of simulated
AGNs that span a wide range in \mbh, \lledd, and S/N, as described in
above.  Our procedure is the following.  We step between the minimum
and maximum luminosity in steps of 0.5 dex (a factor of 3 in
luminosity), and recalculate the total magnitude and thus typical S/N
of the source.  Using the measured \mbh\ and the new \lha, we also
assign an Eddington ratio to each source.  All simulations with
similar \mbh, \lledd, and S/N are then selected, and a weighted-mean
detection fraction is calculated (that is, we assign more weight to
the objects with the most comparable values in each parameter).  The
benefit of this approach is that the mass-dependent biases highlighted
above are built into the calculation.  We are assuming, however, that
the galaxy level scales with the BH mass.  If this assumption is
severely violated, then the galaxy light may overwhelm the AGN light
more rapidly than assumed in these simulations.  

Using the calculated minimum and maximum luminosities and detection
fractions, we derive the maximum likelihood luminosity function shown
in Figure 5.  That we find such good agreement using two different
methodologies is very encouraging.  As an additional test, we divide
our sample into redshift bins and verify that our maximum likelihood
luminosity function, combined with the detection fractions calculated
above, accurately predicts the observed distribution of \lha\ in each
bin (Sandage \etal\ 1979).  Following Hao \etal, we
calculate, in each redshift bin, the predicted number of sources per
\lha\ bin ($F_{\rm p}$) as:
\begin{equation}
F_{\rm p}(L)\Delta L = \sum_i \frac{p_i(L) \Phi(L) \Delta L}
{\int p_i(L\arcmin) \Phi(L\arcmin) dL\arcmin}.
\end{equation}
We then compare the calculated $F_{\rm p}$ with the observed \lha\
distribution, as shown in Figure B13.  As can be seen, the agreement
is very satisfactory, suggesting that our luminosity function provides
a good description of the data.

%%%%%%%%%%%%%%%%%%%%%%%%%%%%%%%%%%%%%%%%%%%%%%%%%%%%%%%%%%%%%%%%%%%%
%%BoundingBox: 
%\begin{figure*}
%\vbox{ 
%\vskip -0.1truein
\hskip 0.5in
\psfig{file=testz.epsi,width=0.5\textwidth,keepaspectratio=true,angle=-90}
\vskip -0mm
\figcaption[]{
A comparison between the actual distribution of \lha\ (crosses and 
dotted lines) and that predicted by 
the luminosity function and detection fractions (solid histograms).
Each panel shows a different redshift bin.  The good agreement provides 
an important sanity check on our luminosity function.
\label{testz}}
%\end{figure*}
\vskip 5mm
%%%%%%%%%%%%%%%%%%%%%%%%%%%%%%%%%%%%%%%%%%%%%%%%%%%%%%%%%%%%%%%%%%%%%
%\noindent

\end{document}